\documentclass[%
 reprint,
 amsmath,amssymb,
prd,
]{revtex4-2}

\usepackage{graphicx}
\usepackage{dcolumn}
\usepackage{bm}
\usepackage{hyperref}
\usepackage{enumitem}

\def\Xmax{$X_\text{max}$\phantom{.}}
\def\XmaxFULLSTOP{$X_\text{max}$}
\def\Ymax{$Y_\text{max}$\phantom{.}}

\begin{document}

\preprint{APS/123-QED}

\title{Radio Measurements of the Depth of Air-Shower Maximum at the Pierre Auger Observatory}%
\author{
A.~Abdul Halim$^{13}$,
P.~Abreu$^{73}$,
M.~Aglietta$^{55,53}$,
I.~Allekotte$^{1}$,
K.~Almeida Cheminant$^{71}$,
A.~Almela$^{7,12}$,
R.~Aloisio$^{46,47}$,
J.~Alvarez-Mu\~niz$^{79}$,
J.~Ammerman Yebra$^{79}$,
G.A.~Anastasi$^{55,53}$
% ...
}
\affiliation{}
\collaboration{Pierre Auger Collaboration}

\author{\phantom{1}}
\affiliation{
\begin{description}[labelsep=0.2em,align=right,labelwidth=0.7em,labelindent=0em,leftmargin=2em,noitemsep]
\item[$^{1}$] Centro At\'omico Bariloche and Instituto Balseiro (CNEA-UNCuyo-CONICET), San Carlos de Bariloche, Argentina
\item[$^{2}$] Departamento de F\'\i{}sica and Departamento de Ciencias de la Atm\'osfera y los Oc\'eanos, FCEyN, Universidad de Buenos Aires and CONICET, Buenos Aires, Argentina
\item[$^{3}$] IFLP, Universidad Nacional de La Plata and CONICET, La Plata, Argentina
\item[$^{4}$] Instituto de Astronom\'\i{}a y F\'\i{}sica del Espacio (IAFE, CONICET-UBA), Buenos Aires, Argentina
\item[$^{5}$] Instituto de F\'\i{}sica de Rosario (IFIR) -- CONICET/U.N.R.\ and Facultad de Ciencias Bioqu\'\i{}micas y Farmac\'euticas U.N.R., Rosario, Argentina
\item[$^{6}$] Instituto de Tecnolog\'\i{}as en Detecci\'on y Astropart\'\i{}culas (CNEA, CONICET, UNSAM), and Universidad Tecnol\'ogica Nacional -- Facultad Regional Mendoza (CONICET/CNEA), Mendoza, Argentina
\item[$^{7}$] Instituto de Tecnolog\'\i{}as en Detecci\'on y Astropart\'\i{}culas (CNEA, CONICET, UNSAM), Buenos Aires, Argentina
\item[$^{8}$] International Center of Advanced Studies and Instituto de Ciencias F\'\i{}sicas, ECyT-UNSAM and CONICET, Campus Miguelete -- San Mart\'\i{}n, Buenos Aires, Argentina
\item[$^{9}$] Laboratorio Atm\'osfera -- Departamento de Investigaciones en L\'aseres y sus Aplicaciones -- UNIDEF (CITEDEF-CONICET), Argentina
\item[$^{10}$] Observatorio Pierre Auger, Malarg\"ue, Argentina
\item[$^{11}$] Observatorio Pierre Auger and Comisi\'on Nacional de Energ\'\i{}a At\'omica, Malarg\"ue, Argentina
\item[$^{12}$] Universidad Tecnol\'ogica Nacional -- Facultad Regional Buenos Aires, Buenos Aires, Argentina
\item[$^{13}$] University of Adelaide, Adelaide, S.A., Australia
\item[$^{14}$] Universit\'e Libre de Bruxelles (ULB), Brussels, Belgium
\item[$^{15}$] Vrije Universiteit Brussels, Brussels, Belgium
\item[$^{16}$] Centro Brasileiro de Pesquisas Fisicas, Rio de Janeiro, RJ, Brazil
\item[$^{17}$] Centro Federal de Educa\c{c}\~ao Tecnol\'ogica Celso Suckow da Fonseca, Petropolis, Brazil
\item[$^{18}$] Instituto Federal de Educa\c{c}\~ao, Ci\^encia e Tecnologia do Rio de Janeiro (IFRJ), Brazil
\item[$^{19}$] Universidade de S\~ao Paulo, Escola de Engenharia de Lorena, Lorena, SP, Brazil
\item[$^{20}$] Universidade de S\~ao Paulo, Instituto de F\'\i{}sica de S\~ao Carlos, S\~ao Carlos, SP, Brazil
\item[$^{21}$] Universidade de S\~ao Paulo, Instituto de F\'\i{}sica, S\~ao Paulo, SP, Brazil
\item[$^{22}$] Universidade Estadual de Campinas, IFGW, Campinas, SP, Brazil
\item[$^{23}$] Universidade Estadual de Feira de Santana, Feira de Santana, Brazil
\item[$^{24}$] Universidade Federal de Campina Grande, Centro de Ciencias e Tecnologia, Campina Grande, Brazil
\item[$^{25}$] Universidade Federal do ABC, Santo Andr\'e, SP, Brazil
\item[$^{26}$] Universidade Federal do Paran\'a, Setor Palotina, Palotina, Brazil
\item[$^{27}$] Universidade Federal do Rio de Janeiro, Instituto de F\'\i{}sica, Rio de Janeiro, RJ, Brazil
\item[$^{28}$] Universidade Federal do Rio de Janeiro (UFRJ), Observat\'orio do Valongo, Rio de Janeiro, RJ, Brazil
\item[$^{29}$] Universidade Federal Fluminense, EEIMVR, Volta Redonda, RJ, Brazil
\item[$^{30}$] Universidad de Medell\'\i{}n, Medell\'\i{}n, Colombia
\item[$^{31}$] Universidad Industrial de Santander, Bucaramanga, Colombia
\item[$^{32}$] Charles University, Faculty of Mathematics and Physics, Institute of Particle and Nuclear Physics, Prague, Czech Republic
\item[$^{33}$] Institute of Physics of the Czech Academy of Sciences, Prague, Czech Republic
\item[$^{34}$] Palacky University, Olomouc, Czech Republic
\item[$^{35}$] CNRS/IN2P3, IJCLab, Universit\'e Paris-Saclay, Orsay, France
\item[$^{36}$] Laboratoire de Physique Nucl\'eaire et de Hautes Energies (LPNHE), Sorbonne Universit\'e, Universit\'e de Paris, CNRS-IN2P3, Paris, France
\item[$^{37}$] Univ.\ Grenoble Alpes, CNRS, Grenoble Institute of Engineering Univ.\ Grenoble Alpes, LPSC-IN2P3, 38000 Grenoble, France
\item[$^{38}$] Universit\'e Paris-Saclay, CNRS/IN2P3, IJCLab, Orsay, France
\item[$^{39}$] Bergische Universit\"at Wuppertal, Department of Physics, Wuppertal, Germany
\item[$^{40}$] Karlsruhe Institute of Technology (KIT), Institute for Experimental Particle Physics, Karlsruhe, Germany
\item[$^{41}$] Karlsruhe Institute of Technology (KIT), Institut f\"ur Prozessdatenverarbeitung und Elektronik, Karlsruhe, Germany
\item[$^{42}$] Karlsruhe Institute of Technology (KIT), Institute for Astroparticle Physics, Karlsruhe, Germany
\item[$^{43}$] RWTH Aachen University, III.\ Physikalisches Institut A, Aachen, Germany
\item[$^{44}$] Universit\"at Hamburg, II.\ Institut f\"ur Theoretische Physik, Hamburg, Germany
\item[$^{45}$] Universit\"at Siegen, Department Physik -- Experimentelle Teilchenphysik, Siegen, Germany
\item[$^{46}$] Gran Sasso Science Institute, L'Aquila, Italy
\item[$^{47}$] INFN Laboratori Nazionali del Gran Sasso, Assergi (L'Aquila), Italy
\item[$^{48}$] INFN, Sezione di Catania, Catania, Italy
\item[$^{49}$] INFN, Sezione di Lecce, Lecce, Italy
\item[$^{50}$] INFN, Sezione di Milano, Milano, Italy
\item[$^{51}$] INFN, Sezione di Napoli, Napoli, Italy
\item[$^{52}$] INFN, Sezione di Roma ``Tor Vergata'', Roma, Italy
\item[$^{53}$] INFN, Sezione di Torino, Torino, Italy
\item[$^{54}$] Istituto di Astrofisica Spaziale e Fisica Cosmica di Palermo (INAF), Palermo, Italy
\item[$^{55}$] Osservatorio Astrofisico di Torino (INAF), Torino, Italy
\item[$^{56}$] Politecnico di Milano, Dipartimento di Scienze e Tecnologie Aerospaziali , Milano, Italy
\item[$^{57}$] Universit\`a del Salento, Dipartimento di Matematica e Fisica ``E.\ De Giorgi'', Lecce, Italy
\item[$^{58}$] Universit\`a dell'Aquila, Dipartimento di Scienze Fisiche e Chimiche, L'Aquila, Italy
\item[$^{59}$] Universit\`a di Catania, Dipartimento di Fisica e Astronomia ``Ettore Majorana``, Catania, Italy
\item[$^{60}$] Universit\`a di Milano, Dipartimento di Fisica, Milano, Italy
\item[$^{61}$] Universit\`a di Napoli ``Federico II'', Dipartimento di Fisica ``Ettore Pancini'', Napoli, Italy
\item[$^{62}$] Universit\`a di Palermo, Dipartimento di Fisica e Chimica ''E.\ Segr\`e'', Palermo, Italy
\item[$^{63}$] Universit\`a di Roma ``Tor Vergata'', Dipartimento di Fisica, Roma, Italy
\item[$^{64}$] Universit\`a Torino, Dipartimento di Fisica, Torino, Italy
\item[$^{65}$] Benem\'erita Universidad Aut\'onoma de Puebla, Puebla, M\'exico
\item[$^{66}$] Unidad Profesional Interdisciplinaria en Ingenier\'\i{}a y Tecnolog\'\i{}as Avanzadas del Instituto Polit\'ecnico Nacional (UPIITA-IPN), M\'exico, D.F., M\'exico
\item[$^{67}$] Universidad Aut\'onoma de Chiapas, Tuxtla Guti\'errez, Chiapas, M\'exico
\item[$^{68}$] Universidad Michoacana de San Nicol\'as de Hidalgo, Morelia, Michoac\'an, M\'exico
\item[$^{69}$] Universidad Nacional Aut\'onoma de M\'exico, M\'exico, D.F., M\'exico
\item[$^{70}$] Universidad Nacional de San Agustin de Arequipa, Facultad de Ciencias Naturales y Formales, Arequipa, Peru
\item[$^{71}$] Institute of Nuclear Physics PAN, Krakow, Poland
\item[$^{72}$] University of \L{}\'od\'z, Faculty of High-Energy Astrophysics,\L{}\'od\'z, Poland
\item[$^{73}$] Laborat\'orio de Instrumenta\c{c}\~ao e F\'\i{}sica Experimental de Part\'\i{}culas -- LIP and Instituto Superior T\'ecnico -- IST, Universidade de Lisboa -- UL, Lisboa, Portugal
\item[$^{74}$] ``Horia Hulubei'' National Institute for Physics and Nuclear Engineering, Bucharest-Magurele, Romania
\item[$^{75}$] Institute of Space Science, Bucharest-Magurele, Romania
\item[$^{76}$] Center for Astrophysics and Cosmology (CAC), University of Nova Gorica, Nova Gorica, Slovenia
\item[$^{77}$] Experimental Particle Physics Department, J.\ Stefan Institute, Ljubljana, Slovenia
\item[$^{78}$] Universidad de Granada and C.A.F.P.E., Granada, Spain
\item[$^{79}$] Instituto Galego de F\'\i{}sica de Altas Enerx\'\i{}as (IGFAE), Universidade de Santiago de Compostela, Santiago de Compostela, Spain
\item[$^{80}$] IMAPP, Radboud University Nijmegen, Nijmegen, The Netherlands
\item[$^{81}$] Nationaal Instituut voor Kernfysica en Hoge Energie Fysica (NIKHEF), Science Park, Amsterdam, The Netherlands
\item[$^{82}$] Stichting Astronomisch Onderzoek in Nederland (ASTRON), Dwingeloo, The Netherlands
\item[$^{83}$] Universiteit van Amsterdam, Faculty of Science, Amsterdam, The Netherlands
\item[$^{84}$] Case Western Reserve University, Cleveland, OH, USA
\item[$^{85}$] Colorado School of Mines, Golden, CO, USA
\item[$^{86}$] Department of Physics and Astronomy, Lehman College, City University of New York, Bronx, NY, USA
\item[$^{87}$] Michigan Technological University, Houghton, MI, USA
\item[$^{88}$] New York University, New York, NY, USA
\item[$^{89}$] University of Chicago, Enrico Fermi Institute, Chicago, IL, USA
\item[$^{90}$] University of Delaware, Department of Physics and Astronomy, Bartol Research Institute, Newark, DE, USA
\item[$^{91}$] University of Wisconsin-Madison, Department of Physics and WIPAC, Madison, WI, USA
\item[] -----
\item[$^{a}$] Louisiana State University, Baton Rouge, LA, USA
\item[$^{b}$] Institut universitaire de France (IUF), France
\item[$^{c}$] also at University of Bucharest, Physics Department, Bucharest, Romania
\item[$^{d}$] School of Physics and Astronomy, University of Leeds, Leeds, United Kingdom
\item[$^{e}$] now at Agenzia Spaziale Italiana (ASI).\ Via del Politecnico 00133, Roma, Italy
\item[$^{f}$] Fermi National Accelerator Laboratory, Fermilab, Batavia, IL, USA
\item[$^{g}$] now at Graduate School of Science, Osaka Metropolitan University, Osaka, Japan
\item[$^{h}$] now at ECAP, Erlangen, Germany
\item[$^{i}$] Max-Planck-Institut f\"ur Radioastronomie, Bonn, Germany
\item[$^{j}$] also at Kapteyn Institute, University of Groningen, Groningen, The Netherlands
\item[$^{k}$] Colorado State University, Fort Collins, CO, USA
\item[$^{l}$] Pennsylvania State University, University Park, PA, USA
\end{description}
}%

\date{\today}

\begin{abstract}

The \textit{Auger Engineering Radio Array} (AERA), part of the Pierre Auger Observatory, is currently the largest array of radio antenna stations deployed for the detection of cosmic rays, spanning an area of $17$\,km$^2$ with 153 radio stations. It detects the radio emission of extensive air showers produced by cosmic rays in the $30-80$\,MHz band. Here, we report the AERA measurements of the \textit{depth of the shower maximum} ($X_\text{max}$), a probe for mass composition, at cosmic-ray energies between $10^{17.5}$ to $10^{18.8}$\,eV, which show agreement with earlier measurements with the fluorescence technique at the Pierre Auger Observatory. We show advancements in the method for radio \Xmax reconstruction by comparison to dedicated sets of {\sc CORSIKA}/{\sc CoREAS} air-shower simulations, including steps of reconstruction-bias identification and correction, which is of particular importance for irregular or sparse radio arrays. Using the largest set of radio air-shower measurements to date, we show the radio \Xmax resolution as a function of energy, reaching a resolution better than $15$\,g\,cm$^{-2}$ at the highest energies, demonstrating that radio \Xmax measurements are competitive with the established high-precision fluorescence technique. In addition, we developed a procedure for performing an extensive data-driven study of systematic uncertainties, including the effects of acceptance bias, reconstruction bias, and the investigation of possible residual biases. These results have been cross-checked with air showers measured independently with both the radio and fluorescence techniques, a setup unique to the Pierre Auger Observatory. \end{abstract}

\maketitle


\section{Introduction}\label{sec:introduction}

\begin{figure}[!ht]
\includegraphics[width=\columnwidth]{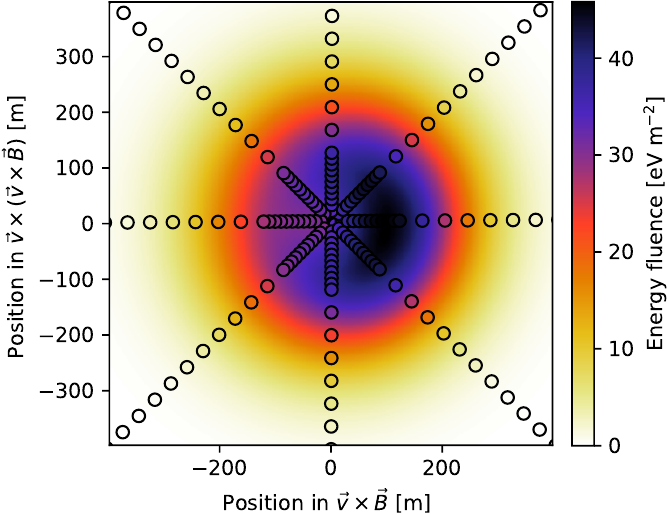}
\caption{\label{fig:footprint2} Example of a footprint of the radio emission on the ground for a simulated air shower with an energy of $8.2\cdot10^{17}$\,eV, a zenith angle of $50.2^\circ$, and a depth of the shower maximum of $749$\,g\,cm$^{-2}$. The strength of the emission is evaluated at simulated antenna positions (markers) and interpolated in between for visibility (background). The footprint has been projected into the \textit{shower plane}, i.e., tilted into the plane perpendicular to the shower axis $\vec{v}$ and rotated to project the magnetic field $\vec{B}$ along the x-axis.}
\end{figure}

\begin{figure}[!ht]
\includegraphics[width=\columnwidth]{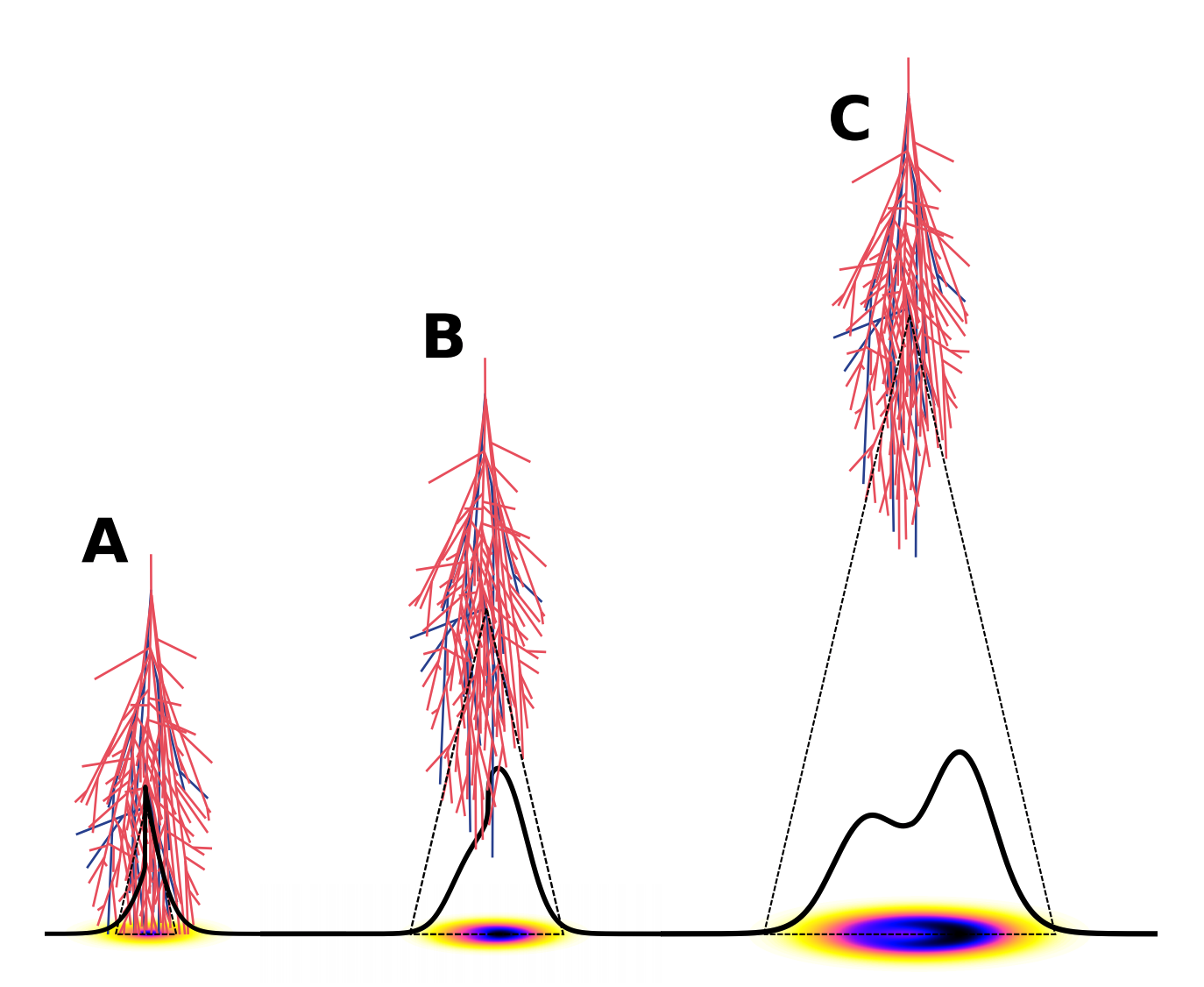}
\caption{\label{fig:footprint1} Schematic view of three air showers that started at different heights in the atmosphere, and their radio emission footprints on the ground. It illustrates that the depth of the shower maximum affects the radio emission footprint both in width and general shape. The asymmetry is a consequence of how the geomagnetic and charge excess radio emission mechanisms interfere during the shower development. This figure has been previously published in~\cite{ref:geoceldf}.}
\end{figure}

The measurement of \textit{ultra-high-energy cosmic rays} (UHECRs) relies on the detection of the products of extensive air showers that are initiated when cosmic rays impact Earth's atmosphere. The study of these air showers allows one to extract their properties and thereby reconstruct important observables, such as the arrival direction of the cosmic-ray primary, its energy, and its particle type. Knowing the particle type is key to understanding the nature and origin of cosmic rays. This is of particular interest in the energy range between $10^{17}$ to $10^{19}$\,eV where the cosmic-ray flux is expected to transition from having Galactic to extragalactic sources (see for example the review~\cite{ref:UHECRWhitePaper}. In the \textit{transition region} a change in mean mass of the primaries, their \textit{mass composition}, could help disentangle source contributions.

The past decades have seen major improvements to the detection of extensive air showers and the reconstruction of air-shower parameters. Though typically this has up until now been the domain of direct particle detection and the observation of air-Cherenkov or fluorescence light, the last two decades also saw the detection of radio emission from air showers coming to maturity~\cite{ref:histlopes,ref:codalema,ref:MGMR,ref:zhaires,ref:MicroCoREAS,ref:2dldf,ref:EnergyScalePRL}. For reviews on the recent progress see e.g.\,~\cite{ref:HuegeReview,ref:Astro2020}. This is important as the radio technique has the advantage of a near-100\% duty cycle and relatively low-cost hardware, while still performing precision measurements of the electromagnetic part of the shower. In the extensive air shower, changing currents, caused by charged particles moving under the influence of the Earth's magnetic field (geomagnetic emission) and by the ionization of the surrounding atmospheric medium (charge excess emission), lead to electromagnetic radiation, predominantly in the MHz to GHz frequency band. Using an array of radio antennas on the ground, the radio emission footprint can then be measured. An example of a simulated radio footprint on the ground is shown in Fig.~\ref{fig:footprint2}. The radio footprint shape depends strongly on particle type and can thus be used to probe the cosmic-ray mass composition. A heavier primary nucleus (which acts roughly as a superposition of multiple lower-energy nucleons) will, on average, interact higher up in the atmosphere and hence produce a wider footprint on the ground than a lighter primary particle. This is illustrated in Fig.~\ref{fig:footprint1}. The particle type itself is not a direct observable, but the atmospheric depth where the shower is maximally developed, the \textit{depth of the shower maximum} $X_\text{max}$, which depends on the particle type, can be related to the shower footprint shape, making \Xmax a probe for the primary particle type. 

Several methods have been used over the past years to reconstruct the particle type from radio signals, most of those relying on determining either the slope, width, or full shape of the \textit{lateral distribution function} (LDF) of the radio footprint to determine $X_\text{max}$~\cite{ref:LOPESresults,ref:Tunka1dLDFXmax,ref:YakutskXmax2019, ref:twodimldf,ref:geoceldf}. In addition, also other methods using for example the slope of the frequency spectrum~\cite{ref:JansenThesis,ref:CanforaThesis} and shape of the shower wavefront have been attempted~\cite{ref:LOPESwavefrontxmax}. Out of all these methods, the highest resolution in \Xmax has been thus far achieved by using the LDF of the radio footprint by fitting of simulated air showers to measured air showers~\cite{ref:LOFARresults0,ref:LOFARresults1}.

In this work, the simulation-fitting method has been further developed for the \textit{Auger Engineering Radio Array} (AERA), by accounting for the effects of the sparse (compared to other radio experiments) and irregular array of radio stations. Also, a thorough investigation of systematic uncertainties has been made. We present the details of the \Xmax reconstruction method and quantify the resolution as a function of cosmic-ray energy. Next, we apply the method to the set of air showers measured by AERA to determine the distributions of \Xmax and interpret this in terms of the cosmic-ray mass composition. We then compare the composition to the results of the \textit{fluorescence detector} (FD) at the Pierre Auger Observatory. Furthermore, we use a subset of air showers, simultaneously measured and independently reconstructed with both AERA and FD to directly evaluate our method and place bounds on the total systematic uncertainty between the two \Xmax detection techniques.

This paper will start with a description of the AERA air-shower reconstruction and the selection of showers in Sec.~\ref{sec:aera}. Then, the \Xmax reconstruction method will be described in Sec.~\ref{sec:xmaxmethod}. In Sec.~\ref{sec:systematics} we make an inventory of systematic uncertainties on the reconstruction of the \Xmax distribution of the selected air showers. The resolution with which \Xmax is reconstructed is then shown in Sec.~\ref{sec:resolution}. Finally, the resulting \Xmax distribution as measured by AERA will be presented in Sec.~\ref{sec:xmaxresults}. 

In an accompanying publication~\cite{ref:AERAXmaxPRL} these results are discussed in the context of the larger field of other measurement techniques and experiments.

\section{AERA Data Reconstruction}\label{sec:aera}

\begin{figure}[t]
\includegraphics[width=\columnwidth]{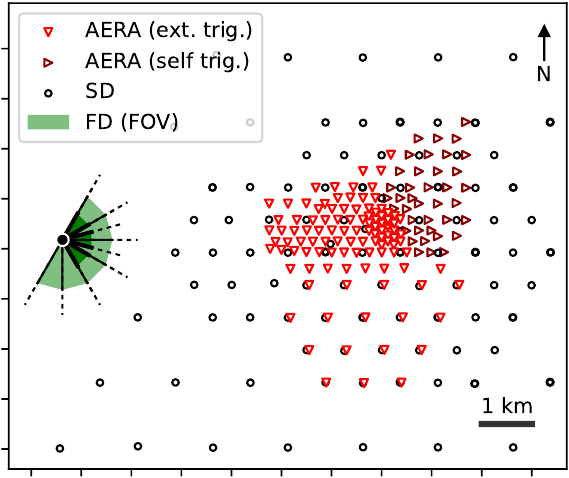}
\caption{\label{fig:AERAmap} Layout of the AERA stations (triangles), marked with whether they are externally- or self-triggered. Also shown is part of the SD array of water-Cherenkov detectors (circles) and the \textit{field of view} (FOV) of one of the FD telescope sites located near AERA. The FD contains $6$ regular telescope bays (light green) and in addition $3$ bays (dark green) looking at higher elevations. Scale and orientation of the layout are indicated with markers.}
\end{figure}

\begin{figure*}[ht]
\includegraphics[width=2\columnwidth]{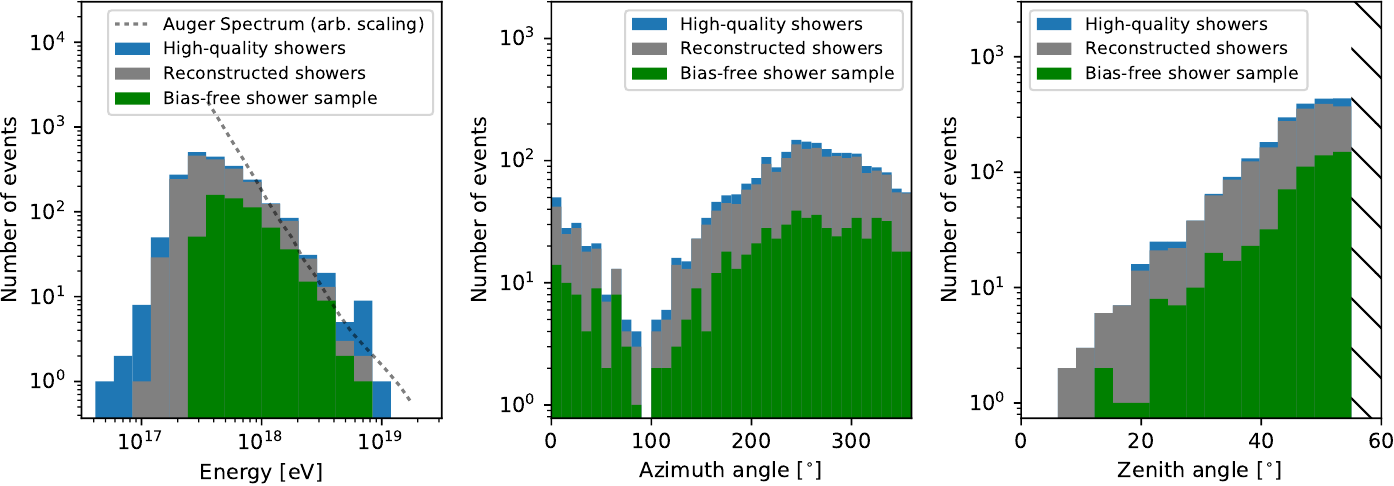}
\caption{\label{fig:Datasel}Distributions of shower energy (left), and the azimuth and zenith angles of the shower arrival direction (center and right, respectively) for the pre-selection of $2153$ high-quality AERA showers (blue), the showers for which \Xmax was reconstructed successfully with our method (gray), and the sample of showers after acceptance and reconstruction cuts are applied (green). The cosmic-ray energy spectrum as measured by Auger~SD~\cite{ref:Infill_Efficiency_v2} (gray dashed line) is scaled to the energy distribution of AERA to illustrate the level of completeness of the AERA event set at the higher energies.}
\end{figure*}

The Pierre Auger Observatory~\cite{ref:augernim} is located near the town of Malarg\"ue in Argentina. It aims at detecting UHECRs up to the highest energies. The observatory covers an area of $3000$\,km$^2$, making it the largest of its kind in the world. The main components of the observatory are an array of $1660$ \textit{water-Cherenkov detectors} (WCDs), also called the \textit{surface detector} (SD), and $27$ fluorescence telescopes (known as the \textit{fluorescence detector}, FD) that overlook the SD. Located near one of the FD sites and within the SD grid is also an array of radio detectors (AERA)~\cite{ref:AERA_Main}. This radio array consists of $153$ autonomous stations, each with two orthogonally placed dipole antennas, that measure the spectrum between $30$ and $80$\,MHz, sampling the signal roughly every $5$\,ns. The measured voltage signals in the antenna arms are converted to an electric field $\vec{E}(t)$ from which we calculate the integrated signal per unit area, conventionally called the \textit{energy fluence} $u$\,[eV/m$^2$]. Part of the measured energy fluence will be from the background noise that will need to be subtracted. One can assume that before the cosmic-ray signal arrives, or long after the cosmic-ray pulse has passed, that the electric field time trace also represents the noise during the time of the signal. The energy fluence can then be calculated as the integral over the time period [$t_{1}$, $t_{N}$] containing the signal, minus the contribution of a pure background time interval [$t_\textup{b,1}$, $t_\textup{b,M}$], where $N$ and $M$ are the respective numbers of samples for the bin size $\Delta t$:
\begin{equation} \label{eq:EnergyFluence}
u = \epsilon_0 c \Delta t \left(  \sum_{i=1}^{N}\left | \vec{E}(t_i) \right |^2 - \frac{t_{N}-t_{1}}{t_\textup{b,M}-t_\textup{b,1}}\sum_{i=1}^{M}\left | \vec{E}(t_\textup{b,i}) \right |^2  \right ),
\end{equation}
where $\epsilon_0$ is the vacuum permittivity and $c$ the speed of light.

The spacing between radio stations varies between $144$\,m and $750$\,m (see Fig.~\ref{fig:AERAmap}) and the array spans a total area of $17$\,km$^2$. While the radio stations can be triggered on the radio signals themselves, in this work we make use of just the external trigger provided by the SD such that we also directly have the measurement of shower energy at our disposal. Consequently, the dark red part of the radio array, shown in  the upper right part in Fig.~\ref{fig:AERAmap}, is a subset of detectors that operate in a self-triggering mode and consequently are not used in this analysis. 

The water-Cherenkov detectors are spaced on a triangular grid of $750$\,m that overlaps with the AERA station grid. Because the radio station spacing is typically much smaller, the estimation of the shower core position and arrival direction is made with the information of cosmic-ray signals in the AERA stations. 

\begin{table}[t]
\caption{\label{tab:datacuts}%
The number of air-shower events remaining after applying the selection criteria sequentially. $\eta$ shows the fraction of showers remaining after each of the cuts. The three sections correspond to the three sets of events in Fig.~\ref{fig:Datasel}.
}
\begin{ruledtabular}
\begin{tabular}{lrr}
\textrm{Quality cut criteria}&
\textrm{Events}&
\textrm{$\eta$[\%]}\\
\colrule
\textit{High-quality shower pre-selection (Sec.~\ref{sec:aera}):}&&\\
 Air-shower candidates                                   & 9336 &    - \\
 SD zenith angle $< 55^\circ$                            & 4874 & 52.2 \\
 SD trigger quality                                      & 2832 & 58.1 \\
 AERA zenith angle $< 55^\circ$                          & 2762 & 97.5 \\
 $\Delta$ AERA and SD arrival direction $< 10^\circ$     & 2733 & 99.0 \\
 No thunderstorm conditions                              & 2160 & 79.0 \\
 $\Delta$ AERA and SD core position $< 400$\,m            & 2153 & 99.7 \\
 
 \hline\textit{Reconstructed air showers (Sec.~\ref{sec:xmaxmethod}):} &&\\
 Insufficient high-SNR sim. radio signals                & 1967 & 91.4 \\
 High-quality \Xmax parabola fit                         & 1725 & 87.7 \\
 Valid \Xmax uncertainty and bias estimation             & 1625 & 94.2 \\
 
 \hline\textit{Bias-free shower sample (Sec.~\ref{sec:systematics}):}  &&\\
 $E\geq10^{17.5}$\,eV                                     & 1107 & 68.1 \\
 Acceptance cut                                          & 594  & 53.7 \\

\end{tabular}
\end{ruledtabular}
\end{table}

From the $7$ years of AERA measurements (2013/04 - 2019/11), we select air-shower candidate events that were triggered by SD, which meet the requirement of a certain quality in terms of clustering of triggered SD stations~\cite[p.~79]{ref:PontThesis}. Additionally, we select on the events where at least $5$ AERA stations have measured a signal with a signal-to-noise ratio above $10$ (the signal is defined here as the square of the maximum of the Hilbert envelope of the electric field and the noise as the square of the RMS of the electric field in a time window away from the signal), a lower limit set by the requirements for the \Xmax reconstruction in Sec.~\ref{sec:xmaxmethod}. As part of these criteria, an algorithm was implemented to reject stations from the shower reconstruction in case of hardware failures or excessive \textit{radio-frequency interference} (RFI) background signals, which is monitored every $100$\,seconds. We also limit the data set to showers arriving from within $55^{\circ}$ of the zenith. Firstly, because the reconstruction at higher inclinations is currently an active field of study~\cite{ref:hassd,ref:hasrd}, and secondly, because sensitivity to \Xmax decreases for higher inclinations since the emission region will be more distant. We require both the SD and AERA arrival direction reconstruction to find angles below $55^{\circ}$ and also require agreement between the arrival direction within $10^{\circ}$ and core position within 400~$m$. This acts only as a rejection of outlier values due to bad reconstructions; the arrival direction and core position reconstruction are much better than this (about $1^\circ$ and $50$\,m, respectively, for SD and similar or smaller for AERA)~\cite{ref:sdarrivalres,res:sdcoreres}. Furthermore, periods of enhanced atmospheric electric field conditions, such as occur during times of thunderstorms, are removed from the data set. To be conservative, events are also rejected if no electric field information was available (accounting for half of the events that are rejected in this step). This results in a pre-selected set of $2153$ showers in the energy range of $10^{17}$ to $10^{19}$\,eV, the lower limit being set by the detection threshold above the radio background level and the upper limit being exposure-limited. In Table~\ref{tab:datacuts} we list these cut criteria and the number of events after each cut. In Fig.~\ref{fig:Datasel} we show the distribution of these air showers as a function of the shower energy (left), and the azimuth and zenith angles of the arrival direction (center and right, respectively). Indicated in blue is the pre-selected set of showers as described above. The gray and green elements in the figure refer to further quality cuts in the reconstruction of \Xmax (gray) and selection of a bias-free sample (green) as will be described in Sec.~\ref{sec:systematics}. To illustrate the completeness of the data set, at least at higher energies, the cosmic-ray flux as measured by the Pierre Auger Observatory~\cite{ref:Infill_Efficiency_v2} has been superimposed and re-scaled to the AERA shower distribution.

Note that the radio signal strength depends on the Lorentz force $F\sim \vec{v}\times \vec{B}$, and thus on the angle between the arrival direction of the shower $\vec{v}$ with respect to the Earth's magnetic field $\vec{B}$, hence the increased suppression of the detected showers as the azimuth angle approached (approximately) $90^\circ$ and the arrival direction becomes more aligned with the magnetic field.

There is a small overlap in the effective field of view of AERA and FD, such that for a subset of $53$ showers in the set of selected showers for AERA also an independent high-quality FD shower reconstruction is available. The number is mainly limited by the distance and different energy dependent apertures of AERA and FD, and the FD duty cycle. We will use these $53$ showers in Sec.~\ref{sec:systematics} for an independent check on the \Xmax reconstruction on an event-by-event basis.

\section{Reconstruction Method for $X_\text{max}$}\label{sec:xmaxmethod}

The method to reconstruct the depth of the shower maximum that we use in this work is based on the method developed for LOFAR~\cite{ref:LOFARresults0,ref:LOFARresults1} where a set of Monte-Carlo (MC) air-shower simulations is generated based on the basic reconstructed properties of a measured air shower such as cosmic-ray energy and arrival direction. The depth of the shower maximum $X_\text{max}$, is affected by shower-to-shower fluctuations and thus similarly varies for each of the simulations. The sensitivity of the radio signals to \Xmax is then used to match the radio signals between measurement and simulations to reconstruct the \Xmax value of the measured air shower. We use the air-shower simulation code {\sc CORSIKA} v7.7100~\cite{ref:Corsikamain} with radio extension {\sc CoREAS}~\cite{ref:MicroCoREAS} and QGSJetII-04~\cite{ref:QGSjetIImain} as our high-energy hadronic interaction model. We include several higher-order effects to simulate the individual measured air showers as precisely as possible. We include the \textit{Global Data Assimilation System} (GDAS) atmospheric model~\cite{ref:gdastool,ref:gdasinoffline} for our time and location-dependent air density and refractive index modelling. A time-variable geomagnetic field model~\cite{ref:magneticmodel} is also used because the majority of the emission being driven by the magnetic field, which changed slightly over the several years of AERA data used in this work. We model the simulated stations to lie on the sloped plane of AERA and add several concentric rings of 'virtual' stations such that we can interpolate the energy fluences with higher precision between AERA station positions. This is done because the core position of the shower is only known to the order of $20$\,m and, when comparing the simulated and measured radio signals, we shift the simulated footprints to correct for the offset caused by this uncertainty.

As input for the shower simulations we use the shower energy from SD. All showers in the data set are triggered by the SD and hence the SD energy measurement is available for each event. For the shower core position and arrival direction, we use the reconstruction from AERA. The stochastic nature of particle interactions in the air shower leads to shower-to-shower fluctuation in \Xmax such that this parameter can be described by a Gumbel distribution~\cite{ref:gumbelmotivation}. We create an ensemble of $27$ simulations for each of the $2153$ selected air showers: $15$ induced by protons and $12$ induced by iron nuclei (intermediate-mass particles are not used as it has been shown to not be necessary~\cite{ref:Xmax_TunkaREX}). We use more proton showers since these cover a larger range in \XmaxFULLSTOP. These primaries and quantities are selected to cover the true distribution of $X_\text{max}$), including the tails of the \Xmax distributions, by varying the initial seeds and height of the first interaction of the primary cosmic ray, while keeping all other input parameters identical. In this way, when comparing the simulated and measured radio signals, we can determine the \Xmax value which best describes the measured signals of the air shower.

\begin{figure}[!ht]
\includegraphics[width=\columnwidth]{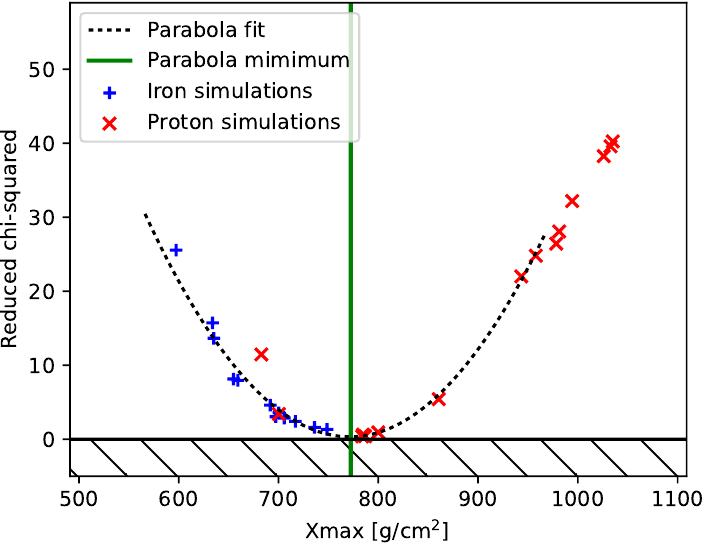}
\caption{\label{fig:Parab} Parabola fit (dashed black line) to the reduced chi-squared values between a measured shower and each of the simulated showers for this event (blue and red markers) [see Eq.~(\ref{eq:XmaxRecChiSquared})] as a function of the true MC \Xmax values for each simulation. The minimum of the parabola (green line) is an estimator for the \Xmax value of the measured shower. This measured shower has an energy of $8.2\cdot10^{17}$\,eV, a zenith angle of $50.2^\circ$, and reconstructed $X_\text{max}=763\pm19$\,g\,cm$^{-2}$. It has been chosen as a representative shower falling in the middle of the AERA energy range (Fig.~\ref{fig:Datasel}, left), being close to the most common zenith angle (Fig.~\ref{fig:Datasel}, right), and having a typical \Xmax resolution (Fig.~\ref{fig:Resolution}).} 
\end{figure}

\begin{figure}[!ht]
    \includegraphics[width=\columnwidth]{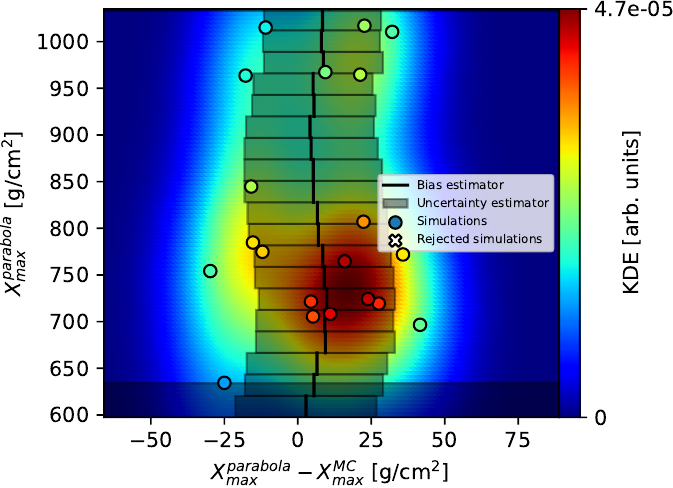}
    \caption{\label{fig:KDE1} The parabola $X_\textup{max}$ values reconstructed for the set of simulations of a single measured air shower (same as Fig~\ref{fig:Parab}), as a function of the deviation to the true MC $X_\textup{max}$ values (dots). A kernel density estimation (background color) is made to estimate the probability density function of the difference at each $X_\textup{max}^\textup{parabola}$ value. From this, a mean $\Delta X_\textup{max,1}^\textup{KDE}$ and width $\delta X_\textup{max,1}^\textup{KDE}$ is derived as first-order estimation of bias an estimation of uncertainty in the $X_\textup{max}^\textup{parabola}$ estimator (gray bands at regular intervals). The shaded band at the bottom illustrates the shallowest \Xmax that can be reconstructed such that the parabola minimum remains contained well within the MC \Xmax range. The bias correction procedure corrects for bias introduced by this restriction.}
    \end{figure}
\begin{figure}[!ht]
    \includegraphics[width=\columnwidth]{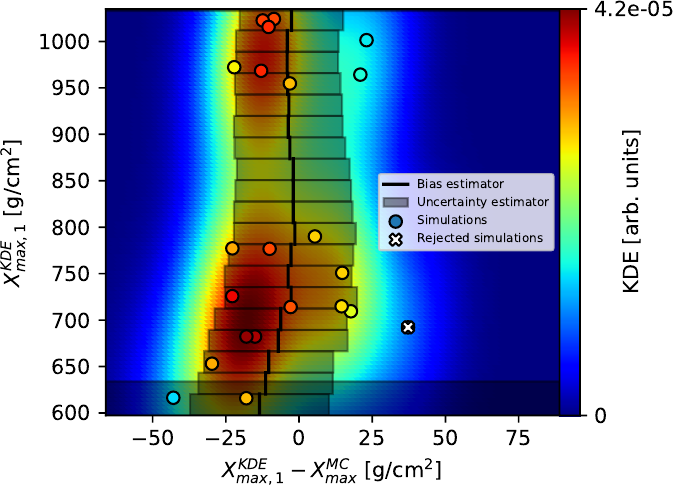}
    \caption{\label{fig:KDE2} Second-order bias correction $\Delta X_\textup{max,2}^\textup{KDE}$ and total uncertainty $\delta X_\textup{max,2}^\textup{KDE}$ after including the effects of a free core and free energy scaling in the minimization procedure. Once again the same event is used as in Fig.~\ref{fig:Parab} and Fig.~\ref{fig:KDE1}.}
\end{figure}

For each shower we quantify the quality of the match between the measured and the simulated radio signals by defining a chi-squared quantity based on the energy fluence $u$ and the corresponding uncertainty $\sigma_u$ of the radio stations:

\begin{equation}
       \chi^2 = \sum_{\textup{stations}} \frac{\left(u_\textup{measured}-S \cdot u_\textup{simulated}(\vec{r}_\textup{shift}) \right )^2}{\sigma_{u_\textup{measured}}^2} \label{eq:XmaxRecChiSquared}.
       \end{equation}
       
The simulated energy fluences $u_\textup{simulated}$ are calculated by applying the AERA antenna response to the pure simulated signals and then reconstructing them as if they were actual measured signals~\cite{ref:RadioOffline} (no noise is added to the simulated signals since we would have to remove it again to calculate the energy fluence, as in Eq.~\ref{eq:EnergyFluence}, needlessly reducing precision. The uncertainty on \Xmax reconstruction due to noise is accounted for later in this section). In the chi-squared measure, we account for the possible systematic uncertainties from the air-shower simulations and the uncertainty on the reconstruction of the shower energy by introducing a scaling parameter $S$ between measured and simulated energy fluences. We also account for the uncertainty on the reconstruction of the shower core position with a core shifting parameter $\vec{r}_{\textup{shift}}$. Suitable starting values for the core shift are taken from either an initial fitting procedure~\cite{ref:geoceldf} or a barycenter calculation. Both free parameters are shared between all simulations for the event under consideration (because a measured shower and its corresponding simulations have just a single core offset and energy scaling between them). 

The chi-squared values for each of the shower simulations as a function of the true MC \Xmax values of those shower simulations can be fitted (locally) with a parabola function as is illustrated in Fig.~\ref{fig:Parab} such that the \Xmax value at the minimum of the parabola fit $X_\text{max}^\text{parabola}$ acts as an estimator for the \Xmax value of the measured shower. The minimum is found by an iterative procedure where the free parameter space of $S$ and $\vec{r}_{\textup{shift}}$ is searched for a global minimum in $\chi^2$. Checks are built into the procedure such that the minimum is in fact a global minimum (using a basin-hopping minimizer~\cite{ref:scipy} and an additional coarse full-parameter space search), and that the parabola fit is well-behaved. We test the validity of this procedure by evaluating this with the reconstruction of each of the simulated showers under realistic ambient noise conditions (from periodic noise measurements with our stations), by leaving out that specific simulation and then minimizing $\chi^2$ using the other $26$ simulations belonging to that particular air-shower event. The minimization then provides $S$ and $\vec{r}_\text{shift}$ parameters for evaluation. We reconstruct $S=1$ within an offset of ($0.9\pm0.4$)\% and a spread of ($23.4\pm0.1$)\%. The bias is negligibly small and the uncertainty is primarily driven by the propagation of the uncertainty on the radio signal itself. The free parameter for the core shift we determine to have a minor bias of $(0.4\pm0.2)$\,m and the spread is found to be ($20.6\pm0.2)$\,m, which is on the same order as the core position resolution of AERA. Hence, the minimization algorithm used to determine the best-fit between measured (or the simulated ones mimicking real measurements) and simulated showers does not introduce any additional biases in the free parameters.

The resolution and possible bias of the parabola-\Xmax reconstruction procedure is evaluated by reconstructing the $X_\text{max}^\text{parabola}$ values of each of the 27 simulated showers, that we have for each measured shower, and comparing these reconstructions to their true Monte Carlo values $X_\text{max}^\text{MC}$. An example of this procedure is shown in Fig.~\ref{fig:KDE1}. It shows the difference between the $X_\text{max}^\text{parabola}$ and $X_\text{max}^\text{MC}$ values for each of the simulations (points) as a function of $X_\text{max}^\text{parabola}$. Note that simulations with a bad $\chi^2$ probability for the parabola fit are shown as rejected (crosses, see Fig.~\ref{fig:KDE2}) and simulations that failed to reconstruct are not shown (the resulting effect on the detector sensitivity to \Xmax is quantified in Sec.~\ref{sec:systematics}). The spread along the horizontal axis is not necessarily a constant value for any $X_\text{max}^\text{parabola}$ and, in addition, there can be a bias that depends on $X_\text{max}^\text{parabola}$ itself. The main reason for this is that the constraining power on \Xmax is determined by the amount and quality of radio signals for a particular air shower and these quantities change with $X_\text{max}$. In addition, the parabola fit in the estimation of $X_\text{max}^\text{parabola}$ will be more difficult to make when the chi-squared minimum is near the edges of the range of $X_\text{max}^\text{MC}$ values. As a consequence, the very low \Xmax values will often be overestimated and the very high values often underestimated. Because of this inherent bias in this estimator, we implement steps to mitigate this. We model the spread and bias of the difference in $X_\text{max}^\text{parabola}$ versus $X_\text{max}^\text{MC}$ by determining the \textit{kernel density estimator} (KDE) for the simulated points (colored background in Fig.~\ref{fig:KDE1}). A KDE is a method to estimate a smooth probability density distribution based on substituting discrete points by smooth functions (Gaussian kernels). We extract from this the mean and $1\sigma$ spread at any desired $X_\text{max}^\text{parabola}$ value (illustrated with regularly spaced black bars). A shift from zero on the horizontal axis then indicates the bias as a function of $X_\text{max}^\text{parabola}$ and the spread of the points provides the uncertainty of the reconstruction. Note that for the spread we have taken into account that the bandwidth of the KDE broadens the spread and we have corrected for this such that the uncertainty on \Xmax that we determine is truly a $1\sigma$ error with respect to the spread in MC \Xmax values.

We perform this procedure in two steps to disentangle, in a more stable way, the effects of the intrinsic uncertainties of our \Xmax reconstruction method and the uncertainties that can arise from the uncertainties on the core position and shower energy (the two free parameters in [Eq.~(\ref{eq:XmaxRecChiSquared})]) that are inherent to just the measured air showers. 

In the first step we fix the shower core position and energy scaling parameters to the true Monte-Carlo values such that we can calculate the KDE (Fig.~\ref{fig:KDE1}) to identify and correct for any bias in the \Xmax estimation 
\begin{equation}
\Delta X_\textup{max,1}^\textup{KDE} \equiv X_\text{max}^\text{parabola}-X_\text{max}^\text{MC} 
\end{equation}
caused by the parabola \Xmax estimation itself. This then provides an improved, \textit{first-order bias-corrected}, estimator for \Xmax 
\begin{equation}
X_\textup{max,1}^\textup{KDE} \equiv X_\text{max}^\text{parabola} - \Delta X_\textup{max,1}^\textup{KDE}.
\end{equation}

In the second step, the \Xmax reconstruction is repeated, but now performed including the $2$ free parameters. In this way, we can separately identify and correct for any \Xmax-reconstruction bias originating from the uncertainties on the measured core position and shower energy that were used as input parameters for the {\sc CORSIKA} simulations. For this we look at the $X_\textup{max,1}^\textup{KDE}$ estimator (i.e., after the first KDE-correction step) for each reconstructed simulation, compare this to the true MC values as before, model it again with a KDE, and as before extract bias and uncertainty estimators. The second-order bias correction is then given by
\begin{equation}
\Delta X_\textup{max,2}^\textup{KDE} \equiv \Delta X_\textup{max,1}^\textup{KDE}-X_\text{max}^\text{MC} 
\end{equation}
By also applying this correction, our final AERA \Xmax estimator 
\begin{equation}
X_\textup{max,2}^\textup{KDE} \equiv X_\textup{max,1}^\textup{KDE} - \Delta X_\textup{max,2}^\textup{KDE}
\end{equation}
is obtained. The spread in the reconstructed \Xmax values in Fig.~\ref{fig:KDE2} provides an estimation of the uncertainty on the \Xmax reconstruction, accounting now for the effects of the full reconstruction procedure as if it was executed on a measured air shower. The spread is extracted from the $1\sigma$ region around the bias estimator value in the KDE model (i.e., the region between the $15.87$\% and $84.13$\% quantiles). For the remainder of this work the estimators for \Xmax and its uncertainty will be called $X_\text{max}$ and $\delta_{X_\text{max}}$, respectively (for the latter $\delta_{X_\text{max}}$ is used instead of $\sigma_{X_\text{max}}$ to avoid confusion with the second moment of the \Xmax distribution, $\sigma(X_\text{max}$), which will be introduced in Sec.~\ref{sec:resolution}).  For both of the steps of the procedure quality-checks have been built into the procedure to guarantee the bias and uncertainty estimators represent the underlying data correctly. In situations where this was not the case, primarily for showers with lower quality signals, events have been rejected because of having an ill-defined bias and uncertainty (see the 'Valid \Xmax uncertainty and bias estimation' cut in Table~\ref{tab:datacuts}. These quality criteria are described in more detail in \cite[p.~163]{ref:PontThesis}).

In the end, this procedure provides an end-to-end estimation of \Xmax uncertainty and bias of the method. However, while the bias correction reduces bias, it can't fully correct it. For example, at the edges of the simulated \Xmax range the KDE is sparsely populated and, hence, there the method only has a partial ability to correct for biases. One could mitigate this further by doing more simulations, but here we were computationally constrained to $27$ simulations per shower. We account for any remaining bias of the reconstruction as systematic uncertainty in Sec.~\ref{sec:systematics}.


\section{Acceptance Cuts and Systematic Uncertainties}\label{sec:systematics}

\begin{figure}[!ht]
\includegraphics[width=\columnwidth]{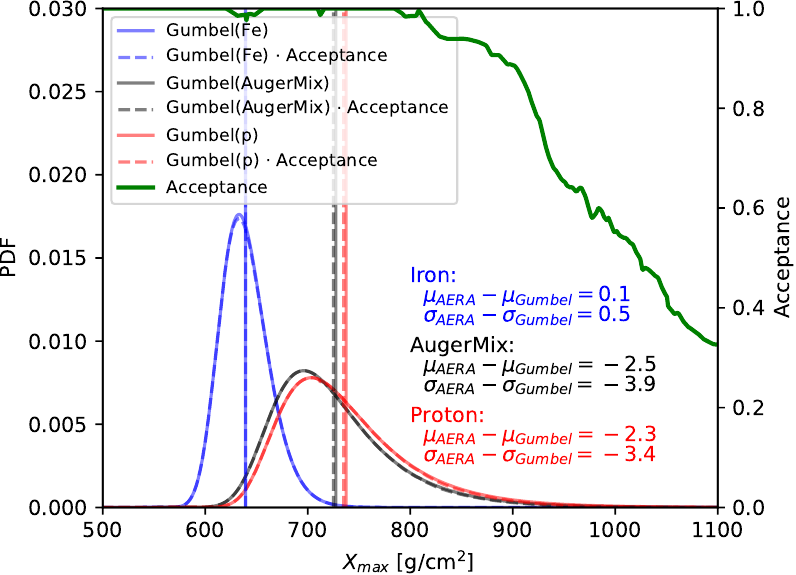}
\caption{\label{fig:AcceptanceSyst} Calculated acceptance for measured AERA showers in the energy bin from $10^{17.95}$ to $10^{18.10}$\,eV (thick green line) and the systematic effect it has on the mean and width of the Gumbel $X_\textup{max}$ distributions (annotated values in g\,cm$^{-2}$) for a pure proton mass composition, pure iron mass composition, and the mixed-mass composition as measured by Auger~FD~\cite{ref:FDXmaxSyst} (solid red, blue, and black lines). The dashed lines are the distributions convolved with the acceptance. The lines are plotted at $50$\% opacity since they match closely. Vertical lines show the respective means. }
\end{figure}

To interpret the distribution of \Xmax we first implement an acceptance cut such that our set of showers is not biased by selection effects. We first apply a cut in energy at $E=10^{17.5}$\,eV, above which the SD trigger we use to read out AERA is fully efficient~\cite{ref:Infill_Efficiency_v2,ref:Infill_Efficiency_v3}. However, not every trigger leads to a high-quality shower in AERA. Hence, we next calculate the detection acceptance for AERA by evaluating the reconstructability of the set of $27$ simulated air showers that were created for each measured shower. We implement the condition that the measured shower should have been detected if it had arrived anywhere within the expected range of \Xmax values as predicted by simulations. Specifically, we require, for any shower we select, that 90\% of the \Xmax values of a Gumbel distribution for both protons and iron nuclei, given the energy of the shower, would be reconstructable by AERA. Removing the events that do not pass the acceptance cut results in $594$ showers. Table~\ref{tab:datacuts} lists these quality cut steps and the final distribution of events can be seen in Fig.~\ref{fig:Datasel} (green shaded area). Fig.~\ref{fig:AcceptanceSyst}, as example, shows the average acceptance (thick green line) for all selected showers with energies between $10^{17.95}$ to $10^{18.10}$\,eV and the average Gumbel distributions for the energies of those showers under the assumption of a composition consisting of just protons (solid red), just iron nuclei (solid blue), and the mixed-mass composition as measured by Auger~FD~\cite{ref:FDXmaxSyst} (Gumbel parametrization for QGSJetII-04~\cite{ref:GumbelMostRecentParam,ref:GumbelMain} are used). At these energies AERA is fully efficient up to about $850$\,g\,cm$^{-2}$, after which the efficiency drops slightly for the tail of the proton Gumbel distribution. For the lowest energies this occurs around $800$\,g\,cm$^{-2}$ (not shown). 

\begin{figure*}[!ht]
\includegraphics[width=2\columnwidth]{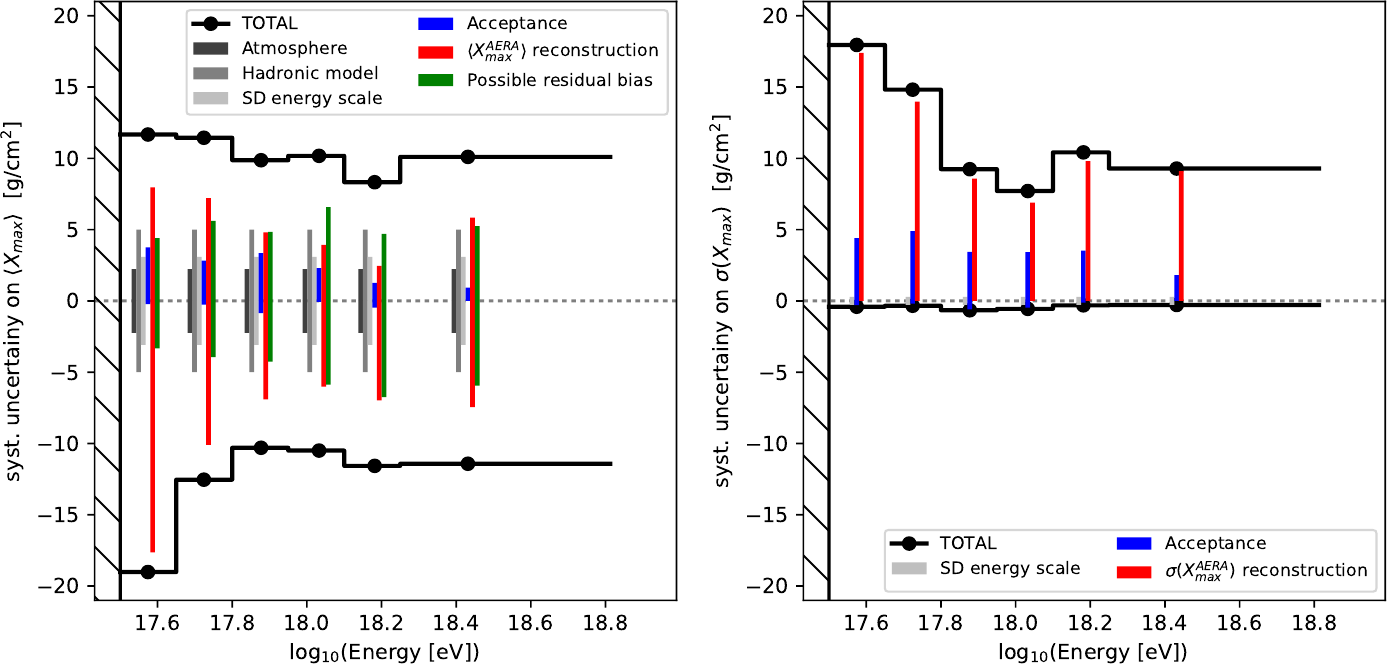}
\caption{\label{fig:TotalSyst}(Left): Overview of upper and lower values of the systematic uncertainties on the mean of the \Xmax distribution ($\langle X_\textup{max} \rangle$). The individual contributions to the total uncertainty are plotted as bars centered in each of the energy bins. The total uncertainty (black lines) is the quadratic sum of the individual contributions. The average energy in each energy bin is shown as black circles. (Right): Overview of systematic uncertainties on the true spread of \Xmax ($\sigma(X_\textup{max})$).}
\end{figure*}

\begin{figure}[!ht]
\includegraphics[width=\columnwidth]{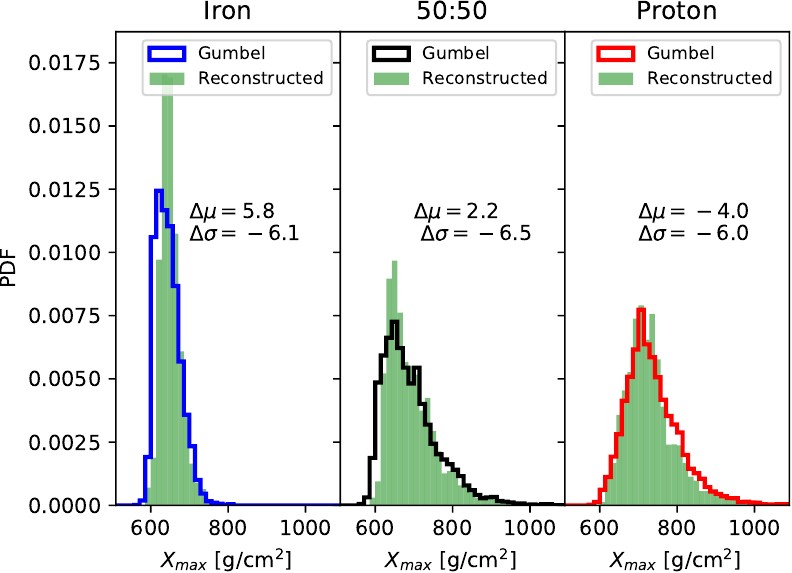}
\caption{\label{fig:MethodSyst} Calculated systematic uncertainties on the $X_\textup{max}$ distribution due to the $X_\textup{max}$ reconstruction method in the energy bin from $10^{17.95}$ to $10^{18.10}$\,eV for a pure proton mass composition (left), pure iron nuclei mass composition (right), and a 50:50 mix of the two (center). The Gumbel distributions (solid line) and how AERA would reconstruct this (green area) are plotted. The Gumbel distributions have been convolved with the AERA measurement uncertainties to allow for direct comparison. The difference between the AERA and Gumbel distributions provides an estimation for the systematic bias on the mean and width of the \Xmax distribution (annotated values in g\,cm$^{-2}$).}
\end{figure}

Although the acceptance is only reduced for extreme \Xmax values we investigate the systematic uncertainty on the mean and width of our measured \Xmax distribution that this would cause. For this, we calculate the effect of the acceptance curve on the Gumbel distributions (dashed lines in Fig.~\ref{fig:AcceptanceSyst}), which are shown to be modified by less
than a few g\,cm$^{-2}$ compared to the solid lines. The resulting differences in the two moments of the distributions are shown as insets in the figure. This calculation is performed for all energy bins and results in a bias of under $4$\,g\,cm$^{-2}$ and $5$\,g\,cm$^{-2}$ on the mean and width, respectively, when assuming the least favourable composition conditions (see blue bars in Fig.~\ref{fig:TotalSyst}). The calculation is included in Appendix~\ref{app:systacceptance}.

Next, we also evaluate the bias the \Xmax reconstruction of individual showers has on the \Xmax distribution. While Sec.~\ref{sec:xmaxmethod} implemented steps to remove \Xmax bias, this is not guaranteed to be sufficient, especially for the deepest and shallowest showers, as explained in that section. Hence, the overall effect on the selected set of air showers is evaluated by reconstructing \Xmax for the air-shower simulations, for which the true MC \Xmax is known, and calculating the effect the reconstruction would have in the case nature would give us a Gumbel \Xmax distribution for protons, iron nuclei, or a 50:50 mix of the two. For the mean of the \Xmax distribution the proton and iron nuclei cases would represent the two extreme cases, since bias occurs mostly for the deepest and shallowest showers. The width of the distribution would be most affected by a mix of proton and iron, hence we evaluate also the case of a 50:50 mix. Fig.~\ref{fig:MethodSyst} shows, as example, the effect on the distribution for the showers in the energy range of $10^{17.95}$ to $10^{18.10}$\,eV. The bias in the width and mean of the distributions is taken as systematic uncertainty, again, to be conservative, under the assumption of the composition with the largest bias. We show the results of this as a function of energy in Fig.~\ref{fig:TotalSyst} (red bars) and show the calculation in Appendix~\ref{app:systmethod}.

We furthermore account for systematic uncertainty of the use of the GDAS atmospheric model~\cite{ref:LOFARresults0,ref:LOFARAtmosStudy} and the choice of hadronic interaction model in the {\sc CORSIKA} simulation code~\cite{ref:LOFARresults0} (these LOFAR result are also valid for AERA due to the similarities of the implementation of GDAS and \Xmax reconstruction methods). An additional systematic uncertainty on the width and mean of the \Xmax distribution at a certain energy arises from the systematic uncertainty in the energy scale~\cite{ref:SD_energyscale}. These effects, shown in Fig.~\ref{fig:TotalSyst}, are all relatively small compared to the uncertainty from the \Xmax reconstruction itself.

Finally, we investigate any possible residual bias remaining in the data set. We check the mean of the \Xmax distribution as a function of geometry-sensitive parameters such as the shower core position and the arrival direction, which by themselves should not cause any trends in the mean \Xmax if there is no residual bias. Because the number of showers for some energy bins is rather limited we combine all showers regardless of energy and correct for the trend in energy. We define \Ymax as the depth of the shower maximum where the elongation rate, the natural increase of the average \Xmax as a function of energy, and change in \Xmax from composition changes with energy has been corrected by subtracting the mean \Xmax of the showers in the respective energy bin $\langle X_\textup{max}^{\Delta E} \rangle$ and normalized to the all-data mean $\langle X_\textup{max} \rangle$:
\begin{equation}
    Y_\textup{max} \equiv X_\textup{max} - (\left\langle X_\textup{max} \right\rangle_{\Delta E} - \left\langle X_\textup{max} \right\rangle). \label{eq:ymax}
    \end{equation}

This now normalizes all values to roughly the average energy of the set of AERA showers and any residual trends in \Ymax with geometry parameters can be investigated on the full set of data.

\begin{figure}[!ht]
\includegraphics[width=\columnwidth]{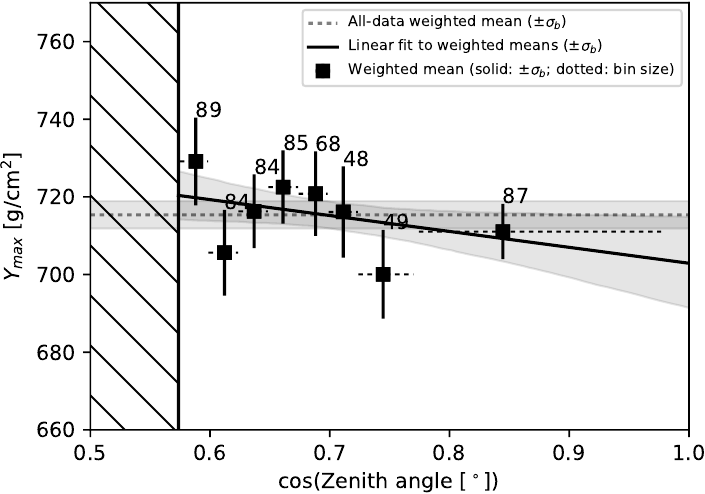}
\caption{\label{fig:ResidualSystZen} Relation between \Ymax [Eq.~(\ref{eq:ymax})] and the cosine of the zenith angle. The mean of $Y_\text{max}$ is shown in equally-spaced bins, or merged bins if containing less than $40$ showers (black squares). The number of events per bin is quoted next to each bin. The solid-line error bars show the uncertainties on the means, determined with bootstrap resampling. The dashed-line bars indicate the extent of each bin. Also shown are the mean of the entire data set (dashed line with $1\sigma$-confidence band) and a linear fit to the mean values (solid line with $1\sigma$-confidence band).}
\end{figure}

Fig.~\ref{fig:ResidualSystZen} shows the effect as a function of the cosine of the shower zenith angle $\theta$. We bin the \Ymax data in equally-sized bins and fit a line to the mean values of the bins. The resulting linear trend (solid line) is shown to be compatible with zero slope within the $1\sigma$ uncertainty band (shaded region) and hence shows no indication of a systematic bias. Possible trends in the azimuth angle $\phi$, the \textit{geomagnetic angle} $\alpha$ (the angle between shower arrival direction and the geomagnetic field, which determines the strength of the geomagnetic emission), and shower core position are also investigated and show similarly small trends compatible with zero slope. Nonetheless, a possible residual bias within these uncertainties can not be excluded. Hence, for each of these geometry parameters we evaluate the effect these possible trends would have on the shower \Xmax values and calculate the magnitude of these possible residual biases in each of the energy bins and for each of the geometry parameters. This procedure is further described in Appendix~\ref{app:residualbias}. 

These parameters are heavily correlated, so their contributions are not added in quadrature, but instead, the extrema are used. This then results in a lower and upper limit on the possible $\langle X_\textup{max} \rangle$ systematic bias of between $-6.8$ and $+6.6$\,g\,cm$^{-2}$, varying slightly depending on energy (see green bars in Fig.~\ref{fig:TotalSyst} and tabulated values in Table~\ref{tab:resbias}). It should be noted that the constraints on these uncertainties are governed by the statistical uncertainty given by the number of showers we have available in this check. The possible systematic biases are well within the statistical uncertainties of $\langle X_\textup{max} \rangle$, so there is no hint that this is a significant bias and thus it should be considered an upper limit on the possible geometry-dependent bias. 

\begin{figure}[!ht]
\includegraphics[width=\columnwidth]{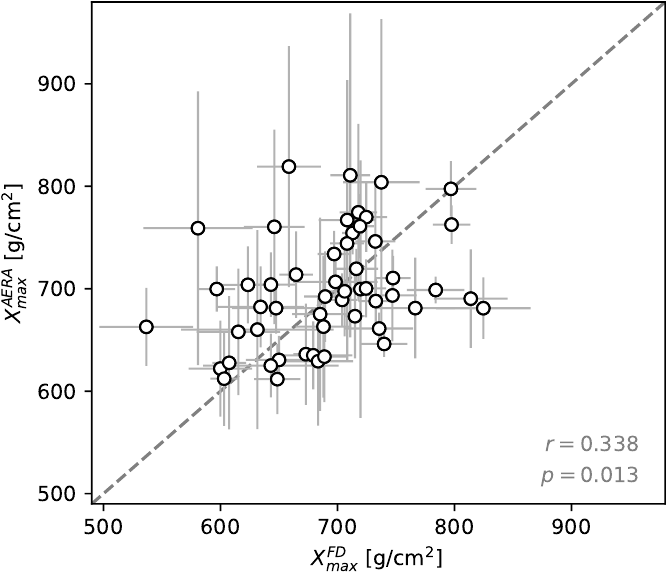}
\caption{\label{fig:FDresultsScatter}Comparison of $X_\textup{max}$ for $53$ showers measured with both FD and AERA. A diagonal (dashed gray line) is plotted to guide the eye. Also shown is the Pearson correlation coefficient $r$ of this data and the corresponding $p$-value (the probability of obtaining an $r$ from uncorrelated data that is at least as high). 
}
\end{figure}

\begin{figure}[!ht]
\includegraphics[width=\columnwidth]{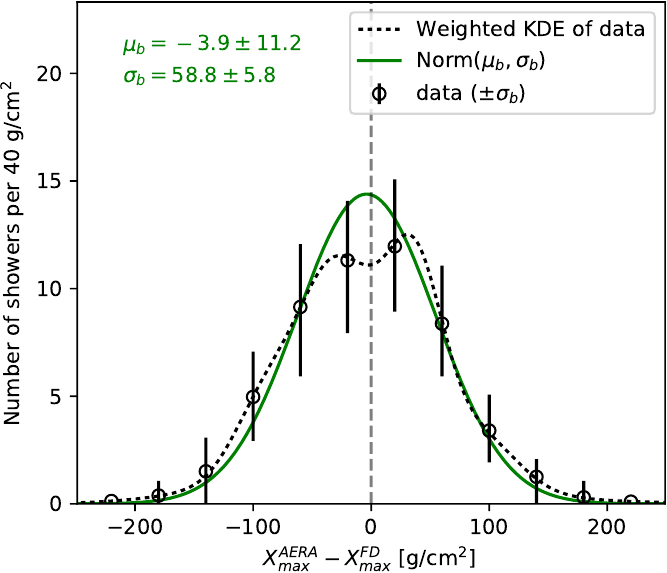}
\caption{\label{fig:FDresults}Results of the comparison of $X_\textup{max}$ for $53$ showers measured with both FD and AERA. Plotted is the weighted KDE (black dashed curve) of the event-to-event differences in $X_\textup{max}$ (sum of $53$ Gaussian distributions with the individual differences as means and combined AERA and FD uncertainty $\delta=(\delta_{X_\text{max}^\textup{AERA}}^2+\delta_{X_\text{max}^\textup{FD}}^2)^{0.5}$ as widths). The black markers show the spread on the KDE, evaluated at intervals of $40$\,g\,cm$^{-2}$, obtained by repeatedly taking $N=53$ samples from the KDE. The calculated weighted mean $\mu_b$ and width $\sigma_b$ of the differences are annotated in the figure (uncertainties are calculated by bootstrap resampling where we repeatedly sample $75$\% of events). For comparison, the Gaussian distribution corresponding to $\mu_b$ and $\sigma_b$ is plotted as solid green curve. Note that the combined resolution of FD and AERA ($53.3\pm5.7$\,g\,cm$^{-2}$, as calculated from the \Xmax uncertainties of the $53$ events) can account for the spread of the difference.}
\end{figure}

It is possible we overestimate our total systematic uncertainty when adding the possible residual bias in quadrature due to correlation with the previously determined uncertainties. Hence, we use the independent \Xmax reconstruction of the fluorescence telescopes, that is available for $53$ air showers in our data set, to obtain an additional and independent estimation on the total systematic uncertainty. The FD data has been prepared as in \cite{ref:FDXmaxSyst} and is shown against the radio reconstruction of \Xmax in Fig.~\ref{fig:FDresultsScatter}. Fig.~\ref{fig:FDresults} shows the distribution of the difference between the two reconstructions to be compatible with zero within $-3.9\pm11.2$\,g\,cm$^{-2}$ for the $53$ showers with energies predominately between $10^{17.5}$ and $10^{18}$\,eV. Taking into account the systematic uncertainty on the FD \Xmax reconstruction itself for these energies (roughly $\pm10$\,g\,cm$^{-2}$)~\cite{ref:FDXmaxSyst} and summing the lower and upper limit in quadrature with the FD uncertainty, results in systematic uncertainty limits of $-18.1$ to $+12.4$\,g\,cm$^{-2}$ on $X_\textup{max}$ for these events, respectively. This estimate for the upper limit of the total systematic uncertainty matches closely to the values for the total systematic uncertainty on $\langle X_\textup{max}\rangle$ of Fig.~\ref{fig:TotalSyst} (on average $-15.6$ and $+11.2$\,g\,cm$^{-2}$, as calculated for the energies of those $53$ events). The combination of these two independent estimations of systematic uncertainties provides further support that our systematic uncertainties are well-understood and that all significant effects have been accounted for. 

Furthermore, the compatibility of the direct event-to-event comparison of the two independent methods hints at the robustness of our understanding of EM cascades in air showers and its implementation in simulations. This is especially important in the context of \Xmax measurements using other aspects of the shower, such as the muonic component~\cite{ref:SDXmax}, which is arguably less well-understood as suggested by the measurements of a significant muon deficit in simulations~\cite{ref:AugerMuonContent}. Our new constraints between the radio and fluorescence \Xmax scales might provide new hints in future studies.


\begin{figure}[t]
\includegraphics[width=\columnwidth]{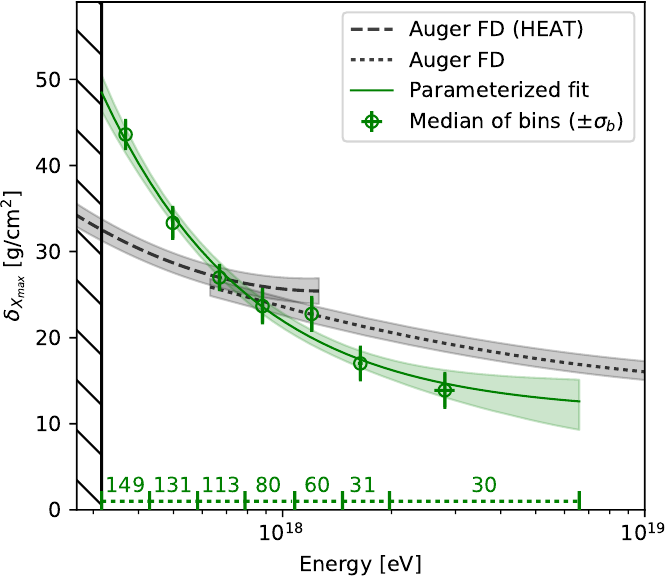}
\caption{\label{fig:Resolution} Resolution of the $X_\textup{max}$ reconstruction method, $\delta_{X_\text{max}}$, as a function of energy in units of column density. Shown per energy bin are the median values of the uncertainties on \Xmax (circles with uncertainties $\sigma_b$ from bootstrap resampling) for all showers in the bias-free sample (Table~\ref{tab:datacuts}) and a parametrized fit [Eq.~(\ref{eq:EnergyResolution})] of the resolution in \Xmax (solid line with $1\sigma$-confidence bands). Also shown are the resolutions achieved by the Auger fluorescence telescopes~\cite{ref:FDXmaxSyst}. The black hatched region at low energy indicates the cut on energy applied earlier. The extent of each energy bin, including the number of showers per bin, is inset at the bottom of the figure.} 
\end{figure}

\section{Resolution}\label{sec:resolution}

In Fig.~\ref{fig:Resolution} we show the uncertainty on $X_\textup{max}$ as a function of shower energy, as determined with our method, for the final bias-free selection of $594$ showers (see Sec.~\ref{sec:systematics}). We find that the median resolution in \Xmax shows a clear relation with shower energy, reaching a resolution of better than $15$\,g\,cm$^{-2}$ in the highest energy bin. We parametrize the resolution as
\begin{equation}
\delta_{X_\text{max}} = a \cdot \sqrt{\frac{10^{18} \text{eV}}{E}} \oplus b \cdot \frac{10^{18}\text{eV}}{E} \oplus c, \label{eq:EnergyResolution}
    \end{equation}

inspired by the energy resolution of electromagnetic calorimeters~\cite{ref:CalorimetryHandbook}, but also functioning as a generic expansion in terms of energy. Here, $a=14.0\pm6.8$\,g\,cm$^{-2}$, $b=12.7\pm2.5$\,g\,cm$^{-2}$, and $c=11.2\pm4.7$\,g\,cm$^{-2}$ are fitted free parameters and $\oplus$ indicates the quadratic sum. The constant term $c$ provides an indication of the resolution that might potentially be obtained for AERA with this method at the highest energies (given this parametrization). The change in resolution of \Xmax is dominated by the uncertainty on the measured radio signals and hence becomes less accurate at lower energy. Comparing our resolution to the resolution achieved by FD, we achieve similar values at the highest energies where the FD reaches $15$\,g\,cm$^{-2}$~\cite{ref:FDXmaxSyst}. Furthermore, our method remains competitive down to lower energies where, for example at $E=10^{17.8}$\,eV the FD achieves the same resolution of $25$\,g\,cm$^{-2}$. The most recent results by the LOFAR radio array, where a similar simulation-fitting method is used to determine $X_\text{max}$, report an average resolution of $19$\,g\,cm$^{-2}$ between $10^{16.8}$ and $10^{18.3}$\,eV~\cite{ref:LOFAR2021}. Despite the much denser antenna spacing of LOFAR, AERA achieves similar resolutions considering the respective energy regimes to which the two experiments are sensitive.

\begin{figure*}[!ht]
\includegraphics[width=2\columnwidth]{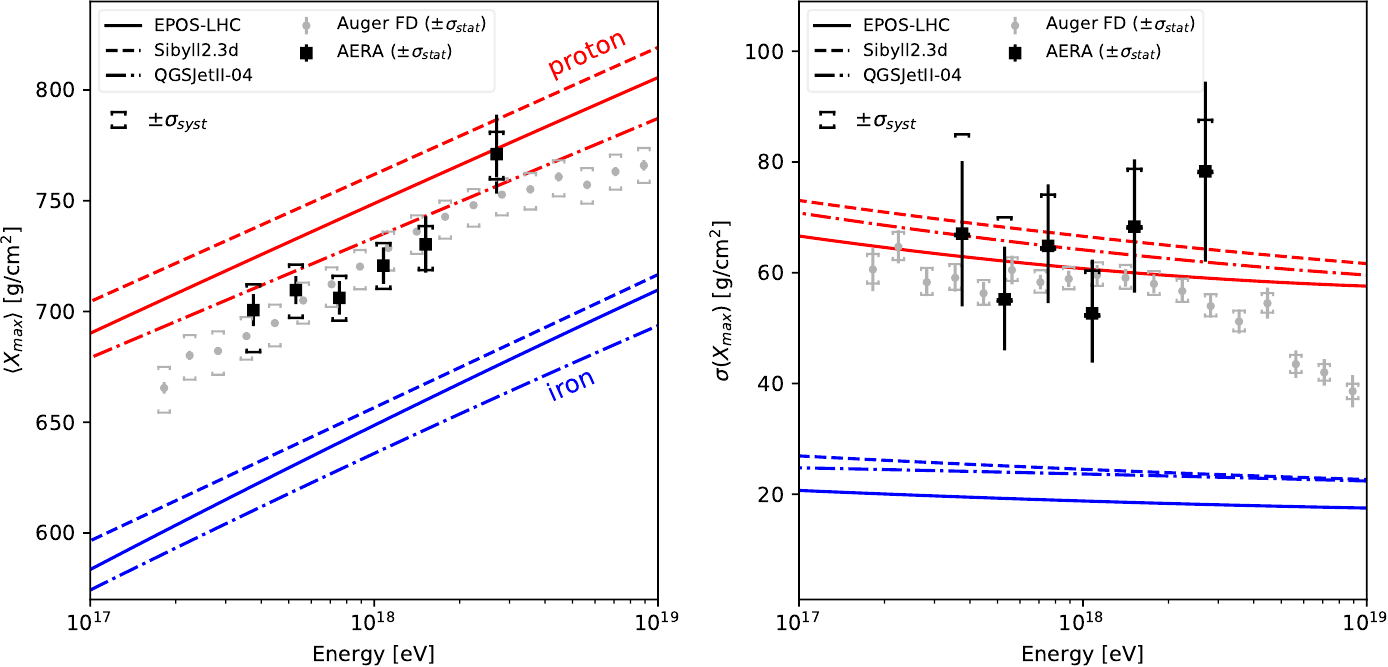}
\caption{\label{fig:elongation}Mean (left) and (resolution-subtracted) width (right) of the $X_\textup{max}$ distribution as measured by AERA in this work (black). The results are compared to predictions from {\sc CONEX} air-shower simulations for three hadronic interaction models (lines) for proton (red) and iron (blue) mass compositions~\cite{ref:hadr_epos,ref:hadr_sibyll,ref:QGSjetIImain,ref:POA_Composition_ICRC2019} and compared to measurements by the FD (gray)~\cite{ref:POA_Composition_ICRC2019}. The statistical uncertainties on the mean and width of the experimental results are plotted as error bars and the systematic uncertainties are shown with caps.}
\end{figure*}

We note that up till now it has been common to quote a single resolution value for \Xmax reconstruction methods for radio experiments, mainly because of a limited number of measured showers being available. Here we show the resolution in \Xmax depends strongly on the shower energy, driven primarily by the strength of radio signals measured in the antennas. Hence, the resolution is a function of detector sensitivity, shower energy, and thus heavily depends on the shower selection criteria. As such, any direct comparison of methods is less straightforward if obtained at sufficiently dissimilar detectors.


\section{The $X_\text{max}$ Moments and the Distribution of $X_\text{max}$}\label{sec:xmaxresults}

\begin{figure*}[!ht]
\includegraphics[width=2\columnwidth]{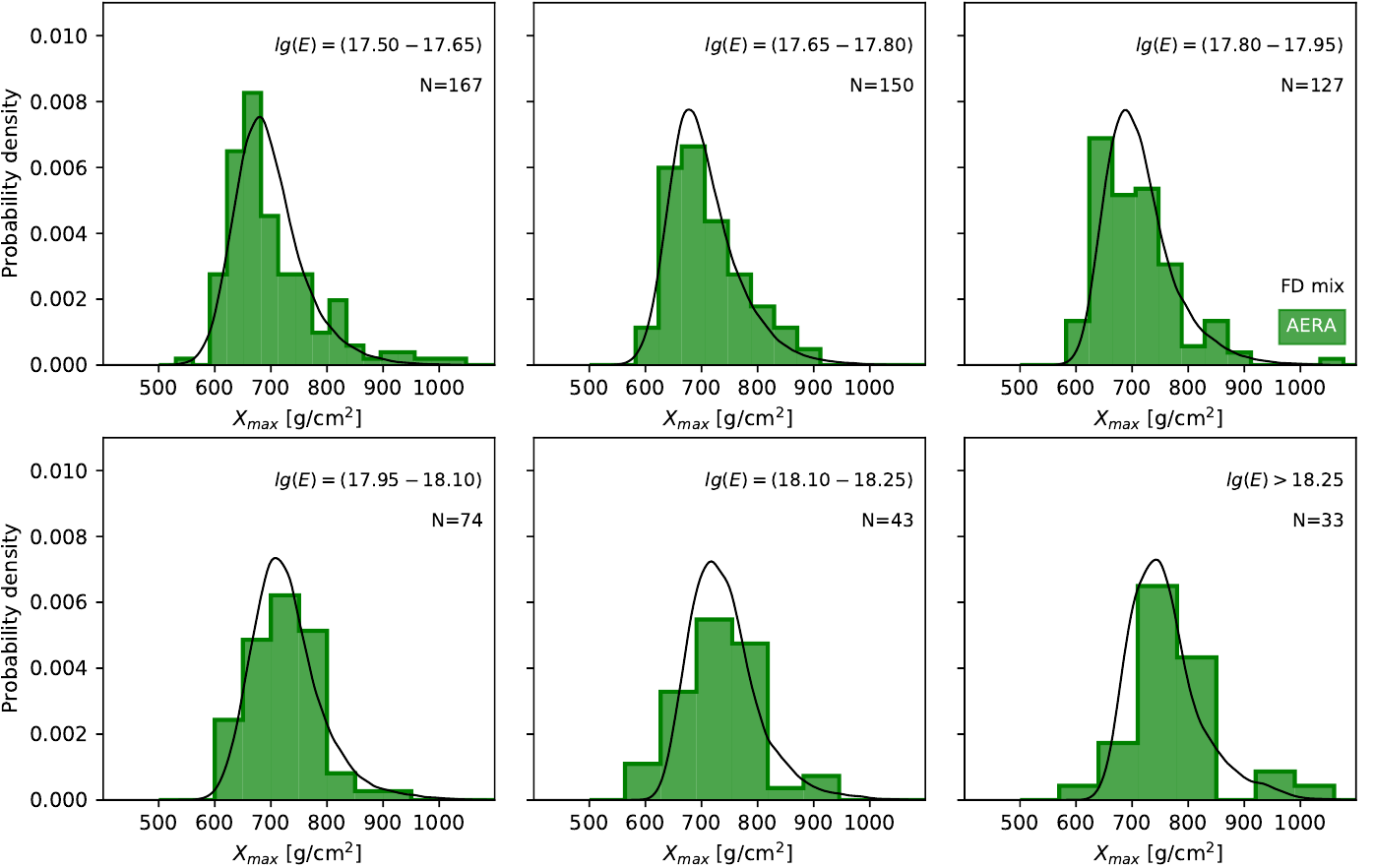}
\caption{\label{fig:XmaxDist}Distribution of \Xmax measured with AERA (green) for six energy bins. These distributions still include the effects of detector resolution, bias, and acceptance. The distributions are compared to the mixed-mass composition as measured by Auger~FD~\cite{ref:FDXmaxSyst} (black), which has been convolved with the detector effects of AERA to allow for direct comparison to the AERA distribution (see main text). Energy ranges and number of showers are annotated in the figures.}
\end{figure*}

From the \Xmax distribution, for each of our six energy bins, we now also calculate the first two moments of the distribution, the mean $\langle X_\textup{max}\rangle$ and the width $\sigma(X_\textup{max})$. To obtain the latter we first subtract in quadrature the width caused by the method uncertainty, such that only the width caused by shower-to-shower fluctuations $\sigma(X_\textup{max})$ remains. The method uncertainty cannot simply be characterized by a single value since the uncertainties on \Xmax for our air showers do not necessarily follow a Gaussian distribution. A bootstrap resampling procedure is applied for this reason and with this we then also calculate the uncertainty on $\sigma(X_\textup{max})$. This procedure is further described in Appendix~\ref{app:sigmaxmax}. The resulting mean and width of the true \Xmax distribution are shown in Fig.~\ref{fig:elongation}, where we also compare this to the results from the FD (gray) and theoretical predictions of three different hadronic interaction models for a mass composition of just protons (red) or just iron nuclei (blue). The systematic uncertainties determined in the previous section are shown with capped markers. Table~\ref{tab:xmaxmoments} in Appendix~\ref{app:tabulated} lists the values for the two moments of the distributions for the six energy bins, together with their statistical and systematic uncertainties.

With these AERA results we show good agreement between the Auger radio and fluorescence measurements of $\langle X_\textup{max} \rangle$, both pointing towards a (mixed)-light composition of cosmic rays at around $E=10^{17.5}$\,eV. Note that the two measurements share the systematic uncertainty on the energy scale, which is constructed from calibration of the SD energy to the FD energy scale~\cite{ref:SD_energyscale}. Taking this contribution out, reduces both systematic uncertainty bands by about $3$\,g\,cm$^{-2}$. Secondly, the determination of the systematic uncertainties due to the reconstruction method for AERA \Xmax data depends on the assumed composition. We have conservatively taken the most pessimistic mass composition scenario for our method. If we were to assume a (mixed) light composition, as the FD reports, then the composition-dependent AERA systematic uncertainty from the reconstruction method and the acceptance calculation, combined, would be reduced to only a small contribution ($\sigma_\text{stat.}^\text{low}=0$\,g\,cm$^{-2}$ at all energies, roughly $\sigma_\text{stat.}^\text{up}=5$\,g\,cm$^{-2}$ for all but the highest energy bin, and roughly $\sigma_\text{stat.}^\text{up}=7$\,g\,cm$^{-2}$ for the highest bin)). This happens because the AERA systematic uncertainties originate primarily from systematic uncertainties on the reconstruction of the very deepest or very shallowest $X_\textup{max}$ values, which would not significantly impact the average \Xmax for a (mixed) light composition. In that case, the total systematic uncertainty on $\langle X_\textup{max}\rangle$ would be just below $\pm10$\,g\,cm$^{-2}$ for all energies.

The AERA results of the second moment of $X_\textup{max}$, shown on the right side of Fig~\ref{fig:elongation} also shows compatibility with the FD results, but have limited resolving power because of the smaller number of showers in comparison to the FD measurements. In the highest AERA energy bin, $\sigma(X_\textup{max})$ is somewhat higher than for the FD. Note that this bin contains just $33$ showers, so a single extreme event in the tail of the \Xmax distribution could result in the observed upwards fluctuation (consequently, the same effect is seen in the first moment for this energy bin). Hence, this single fluctuation is not considered a particularly significant deviation from the FD values within the shown statistical uncertainties.

In Fig.~\ref{fig:XmaxDist} we show the full distributions of the AERA-reconstructed \Xmax values for the bias-free set of $594$ showers (Table~\ref{tab:datacuts}), split again into six energy bins. For comparison, we superimpose the \Xmax distributions for the mixed-mass composition as determined by the Auger~FD measurements (black)~\cite{ref:FDXmaxSyst} which has been convolved with the acceptance, biases, and uncertainties of the AERA \Xmax values and the uncertainty on the FD composition. In this way, a direct comparisons between the FD and AERA \Xmax distribution can be made.

We quantify the compatibility between the FD and AERA \Xmax distributions by calculating the probability we would draw the AERA event sample from the FD distribution, including the known acceptance, uncertainties, and biases from AERA, the uncertainties from the FD composition measurement, and the systematic uncertainties on \Xmax between the FD and AERA. The FD \Xmax distribution is here represented by Gumbel distributions~\cite{ref:GumbelMostRecentParam,ref:GumbelMain} fitted to the FD \Xmax distributions, accounting for detector resolution and acceptance~\cite{ref:FDXmaxSyst}. This allows us to easily sample from the distribution and add AERA measurement effects, such that we can compare quantities between FD and AERA on the same level. 

For each AERA energy bin we repeatedly ($N=1000$) generate FD \Xmax distribution instances from the Gumbel parametrization of the FD composition, evaluated at the energies of the AERA events in that bin. For each instance, we vary the composition within its statistical uncertainties (using many instances of the composition fit to the FD composition~\cite{ref:FDXmaxSyst}). Next, we draw, from each distribution, a set of \Xmax values, with the same number of events as we have for AERA for that energy bin. When drawing values we take into account the AERA acceptance (see e.g., Fig.~\ref{fig:AcceptanceSyst}). Each drawn \Xmax value we shift by its measurement uncertainty and its reconstruction bias. Both quantities are obtained from a parameterization of the difference between our reconstructed \Xmax and the true MC \Xmax for all our {\sc CORSIKA} simulations ($N_\text{sim}=27 \times 594$) as a function of shower energy and MC $X_\textup{max}$. This parametrization contains both the average and spread of the measurement uncertainties and bias, which are then used to draw shifts randomly. With this procedure, we obtain $1000$ 'mock data sets', drawn from the FD distribution, that include the main detection and reconstruction effects of AERA. These now represent \Xmax data set instances that AERA would have measured assuming the FD composition. We can now start to compare these to the actual AERA \Xmax data set to check the compatibility of the AERA and FD distribution.

\begin{figure*}[!htp]
\includegraphics[width=2\columnwidth]{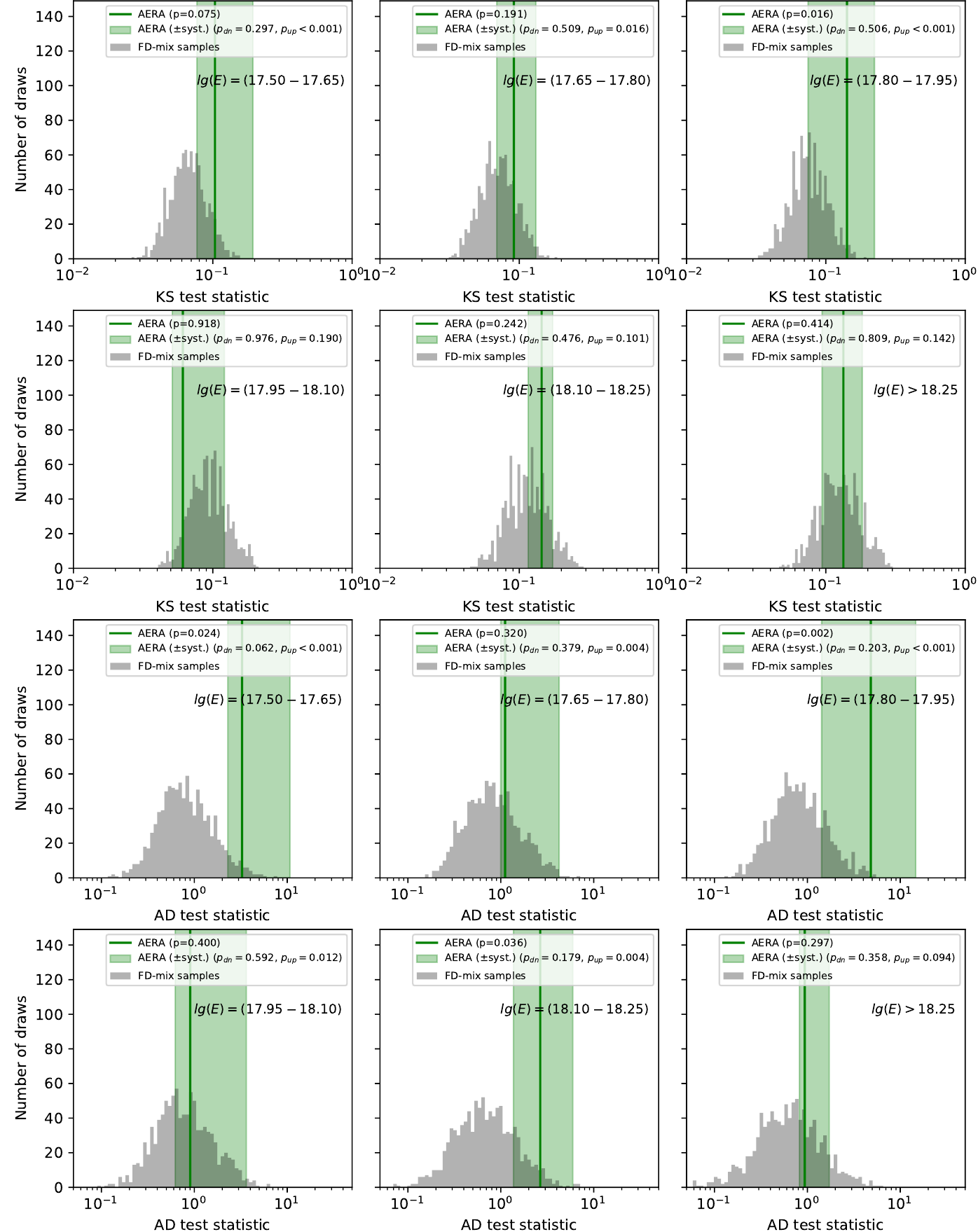}
  \caption{\label{fig:KSADts} Compatibility test of the AERA data and the composition as measured by the Auger FD. Results are shown for six energy bins (values shown in each panel). The top six panels show the distribution of the Kolmogorov-Smirnov (KS) test statistic (gray histograms) for $1000$ \Xmax samples generated from the FD composition as would be measured by AERA (i.e., including the effects of the FD composition uncertainties and AERA reconstruction bias, resolution, and acceptance). The green line in each panel shows the KS test statistic for the AERA data sample (with systematic uncertainty band). The probabilities for compatibility are quoted in the legend and listed in Table~\ref{tab:ts}. The bottom six panels show the same procedure for the Anderson-Darling (AD) test.}
 \end{figure*}

  \begin{table*}[!ht]
\setlength{\extrarowheight}{3pt}
\caption{\label{tab:ts}%
Probabilities for the AERA \Xmax distribution to be drawn from the FD composition, evaluated with the Kolmogorov-Smirnov (KS) and Anderson-Darling (AD) tests, as described in Sec.~\ref{sec:xmaxresults}). Probabilities are quoted per energy bin, listing the values for three scenarios: no shift $\Delta X_\textup{max}$ between the AERA and FD \Xmax distributions (left), the best-matching overall constant shift of $\Delta X_\text{max}^\text{best}=X_\textup{max}^\textup{AERA}-X_\textup{max}^\textup{FD}=-5.5$\,g\,cm$^{-2}$ (center), and best-matching shift for each energy bin and test statistic separately (right). The corresponding shifts are also listed including the range where the test statistic results in $p\geq0.05$.
}
\begin{ruledtabular}
\begin{tabular}{lcc|cc|rcrc}
\multicolumn{1}{c}{}                   & \multicolumn{2}{c|}{no shift}                              & \multicolumn{2}{c|}{constant shift}                        & \multicolumn{4}{c}{variable shift}                                                                                \\ \cline{2-9} 
\multicolumn{1}{c}{lg($E\textup{[eV]}$)} & KS                          & \multicolumn{1}{c|}{AD}      & \multicolumn{1}{c}{KS}      & \multicolumn{1}{c|}{AD}      & \multicolumn{2}{c}{KS}                                  & \multicolumn{2}{c}{AD}                                  \\ \cline{2-9} 
\multicolumn{1}{c}{range}              & \multicolumn{1}{r}{p value} & \multicolumn{1}{r|}{p value} & \multicolumn{1}{r}{p value} & \multicolumn{1}{r|}{p value} & \multicolumn{1}{r}{shift [g\,cm$^{-2}$]} & \multicolumn{1}{c}{p value} & \multicolumn{1}{r}{shift [g\,cm$^{-2}$]} & \multicolumn{1}{c}{p value} \\ \hline
{[}$17.50$, $17.65$)& $0.075$& $0.024$& $0.278$& $0.062$& $  4.5_{ -5.5}^{+ 10.2}$& $0.278$& $  5.1_{ -1.0}^{+  2.5}$& $0.062$\\
{[}$17.65$, $17.80$)& $0.191$& $0.320$& $0.051$& $0.071$& $ -5.0_{ -5.8}^{+ 11.0}$& $0.509$& $ -2.1_{ -7.7}^{+  8.2}$& $0.365$\\
{[}$17.80$, $17.95$)& $0.016$& $0.002$& $0.157$& $0.062$& $  9.1_{ -6.0}^{+ 12.1}$& $0.506$& $ 13.1_{ -8.3}^{+  7.0}$& $0.203$\\
{[}$17.95$, $18.10$)& $0.918$& $0.400$& $0.685$& $0.573$& $  3.0_{-21.5}^{+ 13.4}$& $0.974$& $  5.0_{-13.6}^{+ 12.7}$& $0.566$\\
{[}$18.10$, $18.25$)& $0.242$& $0.036$& $0.476$& $0.111$& $ 10.5_{-26.5}^{+ 19.3}$& $0.476$& $ 16.0_{-14.7}^{+ 12.6}$& $0.179$\\
{[}$18.25$, $\infty$)& $0.414$& $0.297$& $0.200$& $0.172$& $-10.5_{-18.4}^{+ 30.7}$& $0.809$& $ -6.0_{-15.4}^{+ 19.4}$& $0.352$\\
\end{tabular}
\end{ruledtabular}
\end{table*}

To quantify the compatibility we use the \textit{Kolmogorov-Smirnov} (KS) and \textit{Anderson-Darling} (AD) tests (see e.g., \cite{ref:KSADmain}), similarly to the compatibility tests of the \Xmax distributions of the Auger FD and Telescope Array FD~\cite{ref:AugerTAWorkingGroup}. These tests are commonly used to test if a data sample follows a specified distribution. The KS test is particularly suited to test for compatibility in the region around the peak of the distribution, while the AD test also provides sensitivity to the tails of the distribution. Together they are a good measure for the agreement for the overall shape of the distribution. We calculate the KS test statistic $D$ as 
    \begin{equation}
    D = \max_{1\leq i \leq N}\left(F(X_i)-\frac{i-1}{N},\frac{i}{N}-F(X_i)\right),
    \label{eq:KSts}
    \end{equation}
and the AD test statistic $A^2$ as
  \begin{eqnarray}
      A^2 &=& -N-S,\\ \nonumber
      S &=& \sum_{i=1}^{N}\frac{2i-1}{N}\left[ \ln{F(X_i)} + \ln{(1-F(X_{N+1-i})}  \right], \label{eq:ADts}
  \end{eqnarray}
where $X_i$ are the (ordered) data points in a sample, $N$ is the number of samples, and $F$ is the cumulative distribution function of the distribution being tested.

The distribution that we test our samples against we construct as the Gaussian KDE of the $1000$ FD-drawn samples, such that we have a probability density function from which we can obtain the cumulative distribution function $F$. We then calculate $D$ and $A^2$ for each of the individual FD-drawn samples. This provides the range of expected AD and KS test statistics given the sampling effects of AERA measurements. The resulting test statistics are shown in Fig.~\ref{fig:KSADts} (gray histograms). Next, we calculate the test statistics of the AERA data itself and evaluate where it falls within the test statistic distribution (green lines in the same figure). From this, we obtain the probability $p$ of finding a test statistic value larger than the value for the AERA data, i.e., the chance that a sample taken from the distribution under examination is as compatible as the AERA data. The $p$ values are shown in Fig.~\ref{fig:KSADts} and listed in Table~\ref{tab:ts}. We take $p<0.05$ as the threshold to reject the null hypothesis of compatibility. Before interpreting the probabilities, we also have to account for systematic uncertainties. We calculate the effect on the KS and AD test statistics for the AERA data sample for a general shift of the \Xmax distributions allowed within the systematic uncertainties of $\langle X_\textup{max} \rangle$. The systematic uncertainties between AERA and FD consist of the contribution of the FD measurements (roughly $\pm10$\,g\,cm$^{-2}$)~\cite{ref:FDXmaxSyst} and the AERA contributions that were not included in the earlier modelled bias correction when generating $F$, namely the effects of the model for the atmosphere and the hadronic interaction model used in our {\sc CORSIKA} simulations ($\pm5.5$\,g\,cm$^{-2}$ in total, see also Fig~\ref{fig:TotalSyst}). The upper and lower values obtained for the test statistics and the corresponding $p$ values are shown in Fig.~\ref{fig:KSADts} (green bands). 

For the KS test, we find that the AERA data is compatible with the Auger FD composition (FD mix) within the uncertainties for all energy bins, but note that for the third energy bin this requires a small shift allowed within systematic uncertainties. The AD test, similarly, finds compatibility for all energy bins, but requires a shift within systematic uncertainties for the first and third energy bins. We investigate if the required shifts agree with each other, i.e., that compatibility also holds for all energy bins at the same time. For this, we calculate the KS and AD test statistics for the AERA data sample for a range of shifts of \XmaxFULLSTOP. For the simplest scenario of a constant shift at all energies, the best overall match for all energy bins is obtained with a general shift of $\Delta X_\text{max}^\text{best}=X_\textup{max}^\textup{AERA}-X_\textup{max}^\textup{FD}=-5.5\pm0.7$\,g\,cm$^{-2}$ which falls well within the systematic uncertainties. The uncertainty of $0.7$\,g\,cm$^{-2}$ here shows the range of a constant shift where all energy bins show compatibility for both KS and AD test (for the $p\geq0.05$ threshold). The $p$ values for this shift are listed in the central columns of Table~\ref{tab:ts} and show that for both tests $p>0.05$, i.e., compatibility between AERA and FD \Xmax measurements. Furthermore, this shift is in agreement with the shift of $-3.9\pm11.2$\,g\,cm$^{-2}$ obtained for the event-by-event comparison of hybrid events ( Fig~\ref{fig:FDresults}). 

We note that the systematic shift between AERA and FD does not need to be a simple constant, but might depend on energy. Hence, we also calculate the $p$ values for separate shifts per energy bin that lead to the best match between AERA and FD (i.e., highest $p$). The values for $p$ and corresponding shifts (calculated for KS and AD tests separately) are shown in the column on the right in Table~\ref{tab:ts}. We note that the shifts show no significant trend with energy, suggesting that there is no strong dependence on energy in the systematic uncertainties between AERA and FD. Furthermore, the KS and AD tests agree on the obtained shifts, indicating that both the central part and the tails of the \Xmax distribution, respectively, favour such a shift. The uncertainties on the shifts give a good indication that the KS and AD tests are able to constrain the cosmic-ray composition with the AERA measurements. 

In conclusion, the AERA \Xmax distribution provides further support, beyond just the \Xmax moments, for compatibility with the FD \Xmax measurements and suggests a similar general shift between the AERA and FD distributions as for the event-by-event comparison in Fig~\ref{fig:FDresults} (a shift of about $-4$\,g\,cm$^{-2}$). Furthermore, the compatibility indicates that the uncertainties, biases and acceptance of the AERA measurements are well understood.

\section{Conclusions}\label{sec:conclusions}
In this work we show the results of the measurement of the distribution of depth of the shower maximum for air showers measured with the Auger Engineering Radio Array at the Pierre Auger Observatory. We have presented the method used to reconstruct the depth of the shower maximum by comparing measured radio signals to signals from dedicated sets of {\sc CORSIKA}/{\sc CoREAS} air-shower simulations. We show that the resolution of our method is competitive with established techniques to determine $X_\text{max}$. We have selected a set of air showers with minimal selection bias and have quantified any remaining acceptance bias. Furthermore, a detailed study of systematic uncertainties has been conducted accounting for the effects of the reconstruction method, the use of simulation codes, atmospheric models, the energy scale, and possible geometry-dependent residual bias. The total estimated systematic uncertainties on the mean of the \Xmax distribution has been shown to be compatible with an event-by-event comparison of $53$ showers measured by both the Auger fluorescence and radio detectors, indicating a good understanding of the systematic uncertainties in the AERA measurements. In addition, this direct comparison sets a limit on the systematic shift between the AERA and FD \Xmax scale of $-3.9\pm11.2$\,g\,cm$^{-2}$, providing new constraints on our understanding of shower physics. 

The calculated moments of the \Xmax distribution show compatibility with the composition as previously measured by the FD. In addition, the compatibility of the overall shape of the \Xmax distributions between AERA and the FD provides further support beyond the two central moments for the mixed-light composition as previously measured by the FD. Discussions on the comparison of the \Xmax moments to other experiments is available in an accompanying publication~\cite{ref:AERAXmaxPRL}.


\section*{Acknowledgments}

\begin{sloppypar}
The successful installation, commissioning, and operation of the Pierre
Auger Observatory would not have been possible without the strong
commitment and effort from the technical and administrative staff in
Malarg\"ue. We are very grateful to the following agencies and
organizations for financial support:
\end{sloppypar}

\begin{sloppypar}
Argentina -- Comisi\'on Nacional de Energ\'\i{}a At\'omica; Agencia Nacional de
Promoci\'on Cient\'\i{}fica y Tecnol\'ogica (ANPCyT); Consejo Nacional de
Investigaciones Cient\'\i{}ficas y T\'ecnicas (CONICET); Gobierno de la
Provincia de Mendoza; Municipalidad de Malarg\"ue; NDM Holdings and Valle
Las Le\~nas; in gratitude for their continuing cooperation over land
access; Australia -- the Australian Research Council; Belgium -- Fonds
de la Recherche Scientifique (FNRS); Research Foundation Flanders (FWO),
Marie Curie Action of the European Union Grant No.~101107047; Brazil --
Conselho Nacional de Desenvolvimento Cient\'\i{}fico e Tecnol\'ogico (CNPq);
Financiadora de Estudos e Projetos (FINEP); Funda\c{c}\~ao de Amparo \`a
Pesquisa do Estado de Rio de Janeiro (FAPERJ); S\~ao Paulo Research
Foundation (FAPESP) Grants No.~2019/10151-2, No.~2010/07359-6 and
No.~1999/05404-3; Minist\'erio da Ci\^encia, Tecnologia, Inova\c{c}\~oes e
Comunica\c{c}\~oes (MCTIC); Czech Republic -- Grant No.~MSMT CR LTT18004,
LM2015038, LM2018102, LM2023032, CZ.02.1.01/0.0/0.0/16{\textunderscore}013/0001402,
CZ.02.1.01/0.0/0.0/18{\textunderscore}046/0016010 and
CZ.02.1.01/0.0/0.0/17{\textunderscore}049/0008422; France -- Centre de Calcul
IN2P3/CNRS; Centre National de la Recherche Scientifique (CNRS); Conseil
R\'egional Ile-de-France; D\'epartement Physique Nucl\'eaire et Corpusculaire
(PNC-IN2P3/CNRS); D\'epartement Sciences de l'Univers (SDU-INSU/CNRS);
Institut Lagrange de Paris (ILP) Grant No.~LABEX ANR-10-LABX-63 within
the Investissements d'Avenir Programme Grant No.~ANR-11-IDEX-0004-02;
Germany -- Bundesministerium f\"ur Bildung und Forschung (BMBF); Deutsche
Forschungsgemeinschaft (DFG); Finanzministerium Baden-W\"urttemberg;
Helmholtz Alliance for Astroparticle Physics (HAP);
Helmholtz-Gemeinschaft Deutscher Forschungszentren (HGF); Ministerium
f\"ur Kultur und Wissenschaft des Landes Nordrhein-Westfalen; Ministerium
f\"ur Wissenschaft, Forschung und Kunst des Landes Baden-W\"urttemberg;
Italy -- Istituto Nazionale di Fisica Nucleare (INFN); Istituto
Nazionale di Astrofisica (INAF); Ministero dell'Universit\`a e della
Ricerca (MUR); CETEMPS Center of Excellence; Ministero degli Affari
Esteri (MAE), ICSC Centro Nazionale di Ricerca in High Performance
Computing, Big Data and Quantum Computing, funded by European Union
NextGenerationEU, reference code CN{\textunderscore}00000013; M\'exico -- Consejo
Nacional de Ciencia y Tecnolog\'\i{}a (CONACYT) No.~167733; Universidad
Nacional Aut\'onoma de M\'exico (UNAM); PAPIIT DGAPA-UNAM; The Netherlands
-- Ministry of Education, Culture and Science; Netherlands Organisation
for Scientific Research (NWO); Dutch national e-infrastructure with the
support of SURF Cooperative; Poland -- Ministry of Education and
Science, grants No.~DIR/WK/2018/11 and 2022/WK/12; National Science
Centre, grants No.~2016/22/M/ST9/00198, 2016/23/B/ST9/01635,
2020/39/B/ST9/01398, and 2022/45/B/ST9/02163; Portugal -- Portuguese
national funds and FEDER funds within Programa Operacional Factores de
Competitividade through Funda\c{c}\~ao para a Ci\^encia e a Tecnologia
(COMPETE); Romania -- Ministry of Research, Innovation and Digitization,
CNCS-UEFISCDI, contract no.~30N/2023 under Romanian National Core
Program LAPLAS VII, grant no.~PN 23 21 01 02 and project number
PN-III-P1-1.1-TE-2021-0924/TE57/2022, within PNCDI III; Slovenia --
Slovenian Research Agency, grants P1-0031, P1-0385, I0-0033, N1-0111;
Spain -- Ministerio de Econom\'\i{}a, Industria y Competitividad
(FPA2017-85114-P and PID2019-104676GB-C32), Xunta de Galicia (ED431C
2017/07), Junta de Andaluc\'\i{}a (SOMM17/6104/UGR, P18-FR-4314) Feder Funds,
RENATA Red Nacional Tem\'atica de Astropart\'\i{}culas (FPA2015-68783-REDT) and
Mar\'\i{}a de Maeztu Unit of Excellence (MDM-2016-0692); USA -- Department of
Energy, Contracts No.~DE-AC02-07CH11359, No.~DE-FR02-04ER41300,
No.~DE-FG02-99ER41107 and No.~DE-SC0011689; National Science Foundation,
Grant No.~0450696; The Grainger Foundation; Marie Curie-IRSES/EPLANET;
European Particle Physics Latin American Network; and UNESCO.
\end{sloppypar}

\appendix

\section{Calculation of Systematic Uncertainties}

\subsection{Calculation of the Systematic Uncertainty from the Event Selection}\label{app:systacceptance}
    
    The acceptance $A(X_\textup{max})$ for a particular measured air shower was determined from evaluating reconstructability of each of the $27$ air-shower simulations created for that measured shower. The \Xmax values for these showers roughly cover, by design, the range between $500$ and $1100$\,g\,cm$^{-2}$ such that an interpolated acceptance-\Xmax function can be constructed for each shower. The effect the acceptance would have on measuring an \Xmax distribution is quantified by evaluating the two extreme cases of a mass composition described by a \Xmax Gumbel distribution for just protons $G^\textup{p}(X_\textup{max})$, just iron nuclei $G^\textup{Fe}(X_\textup{max})$, a mix 50:50 mix of the two $G^\textup{50:50}(X_\textup{max})$, and the composition as measured by the Auger FD $G^\textup{AugerMix}(X_\textup{max})$. The effect on the mean of such an \Xmax distribution is then calculated and compared to the unaffected Gumbel distribution. The average effect over all events in a particular energy bin is then calculated in order to estimate the systematic shift this might cause on the distribution of measured \Xmax values at these energies. Since the measured composition is not \textit{a priori} known the least favourable composition $C$ (i.e., whichever has the highest bias) is assumed and this is taken as upper and lower limit on the systematic uncertainty on $\langle X_\textup{max}\rangle$:

    \begin{multline}
    \langle X_\textup{max} \rangle_\textup{syst,low}^{A} = \min_\textup{C} \left[ \left\langle \right.\right.\\
    \left.\left.      \left\langle A_\textup{event}(X_\textup{max}) \cdot G_\textup{event}^\textup{C}(X_\textup{max})\right\rangle_{X_\textup{max}}      \right.\right.\\
    \left.\left.      -\left\langle 
      G_\textup{event}^\textup{C}(X_\textup{max})\right\rangle_{X_\textup{max}}\phantom{..} \right.\right.\\    
      \left.\left. \right\rangle_\textup{events}\right],
    \end{multline}
    
    \begin{multline}
    \langle X_\textup{max} \rangle_\textup{syst,up}^{A} = \max_\textup{C} \left[ \left\langle \right.\right.\\
    \left.\left.      \left\langle A_\textup{event}(X_\textup{max}) \cdot G_\textup{event}^\textup{C}(X_\textup{max})\right\rangle_{X_\textup{max}}      \right.\right.\\
    \left.\left.      -\left\langle 
      G_\textup{event}^\textup{C}(X_\textup{max})\right\rangle_{X_\textup{max}}\phantom{..} \right.\right.\\    
      \left.\left. \right\rangle_\textup{events}\right].
    \end{multline}
    
    The same procedure is performed for the width of the \Xmax distribution instead of the mean in order to estimate the systematic uncertainty on $\sigma(X_\textup{max})$. The results, as a function of energy, for both moments of the \Xmax distribution are shown in Fig.~\ref{fig:TotalSyst}.
    
\subsection{Calculation of the Systematic Uncertainty from the \Xmax Reconstruction Method}\label{app:systmethod}

    The systematic effect from any \Xmax reconstruction bias in our method on the mean of the true $X_\textup{max}$ distribution $\langle X_\textup{max}\rangle$ and the spread $\sigma(X_\textup{max})$ are calculated by evaluating the effect on the Gumbel distribution for a pure proton mass composition, pure iron mass composition, and a 50:50 mix of the two. This is done for the same energy bins as for the systematic uncertainty calculation on the acceptance.
    
    The first step is to weigh down the simulations in our simulation set that are rare according to the true \Xmax distributions. These weights are ideally set by the actual distribution in nature, but are \textit{a priori} unknown. What can be assumed, however, is that the composition lies between a pure proton and pure iron mass composition. An upper limit on the systematic uncertainty on \Xmax can then be defined conservatively as the maximum systematic uncertainty determined for the three mass compositions cases. In the same way as was done for the systematic uncertainty on the acceptance. This upper and lower limits will be used as the estimation of the systematic uncertainty on the two moments of the \Xmax distribution.
    
    The weights $w$ for the simulations with $X_\textup{max}^\textup{MC}$ are then defined for a particular composition described by a Gumbel distribution $G$ (or sum of Gumbel distributions for a mixed composition):
    \begin{equation}
    w(X_\textup{max}^\textup{MC}) = \frac{\textup{PDF}(X_\textup{max}^\textup{Gumb})}{\textup{PDF}(X_\textup{max}^\textup{MC})},
    \end{equation}
    where the numerator is the standard Gumbel probability distribution for $X_\textup{max}$ for a particular composition of cosmic rays. The denominator is a Gaussian KDE of the simulated $X_\textup{max}$ values. The latter functions as the probability density function of the simulated values. 

    This then allows us to calculate the mean of the $X_\textup{max}$ distribution from the simulations, under the assumption of a certain composition. The mean of all $X_\textup{max}^\textup{MC}$ (weighted to the Gumbel distribution),
    \begin{equation}
    \langle X_\textup{max}^\textup{MC} \rangle_\textup{Gumb} = \frac{\sum_\textup{sim} \left(X_\textup{max}^\textup{MC} \right )_{\textup{sim}}\cdot w_\textup{sim}}{\sum_\textup{sim} w_\textup{sim}},
    \end{equation}
    by design, approaches the mean of the ideal Gumbel distribution, assuming sufficient number of simulations used in the KDE. For each simulation (with $X_\textup{max}^\textup{MC}$), a value for \Xmax was reconstructed for which the mean can now be calculated in the same way:
    \begin{eqnarray}
    \langle X_\textup{max} \rangle_\textup{Gumb} &=& \frac{\sum_\textup{sim} \left(X_\textup{max} \right )_{\textup{sim}}\cdot w_\textup{sim}}{\sum_\textup{sim} w_\textup{sim}}\\
    &=& \langle X_\textup{max}^\textup{MC} \rangle_\textup{Gumb} + \Delta X_\textup{max}^\textup{bias}.
    \end{eqnarray}

    This provides an estimation of the bias in $\langle X_\textup{max} \rangle$ for a particular composition $C$. As for the calculation of the systematic uncertainty on the acceptance, we conservatively assume the least favourable composition to obtain the systematic uncertainty on the reconstruction method

    \begin{eqnarray}
    \langle X_\textup{max} \rangle_\textup{syst,low}^\textup{method} &=& \max_\textup{C} \left[ \Delta X_\textup{max}^\textup{bias,C}\right]\\
    \langle X_\textup{max} \rangle_\textup{syst,up}^\textup{method} &=& \min_\textup{C} \left[ \Delta X_\textup{max}^\textup{bias,C}\right]
    \end{eqnarray}
    
    A similar calculation is performed to determine the systematic uncertainty on the width of the Gumbel distributions for the AERA values. To obtain this, the mean is replaced by a calculation of the standard deviation, such that $\sigma(X_\textup{max})_\textup{syst,low}^\textup{method}$ and $\sigma(X_\textup{max})_\textup{syst,up}^\textup{method}$ are determined. The resulting systematic uncertainty ranges are shown in Fig.~\ref{fig:TotalSyst}.

\section{Calculation of the possible residual bias}\label{app:residualbias}

    To calculate the effect of possible residual bias depending on geometry parameters such as shower zenith angle or core position, we investigate potential trends with $X_\text{max}$. The mean \Xmax changes with energy, so we defined \Ymax as in Eq.~(\ref{eq:ymax}) such that this expected dependence can be removed and any remaining residual biases can be identified for the data set as a whole.
 
  \begin{figure*}[!ht]
    \includegraphics[width=0.49\textwidth]{ResidualBias_Zen.pdf}
  \hfill
    \includegraphics[width=0.48\textwidth]{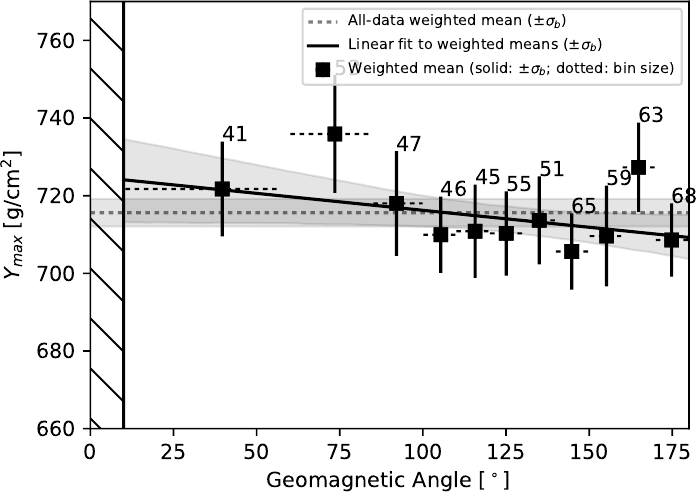}
  \hfill
    \includegraphics[width=0.49\textwidth]{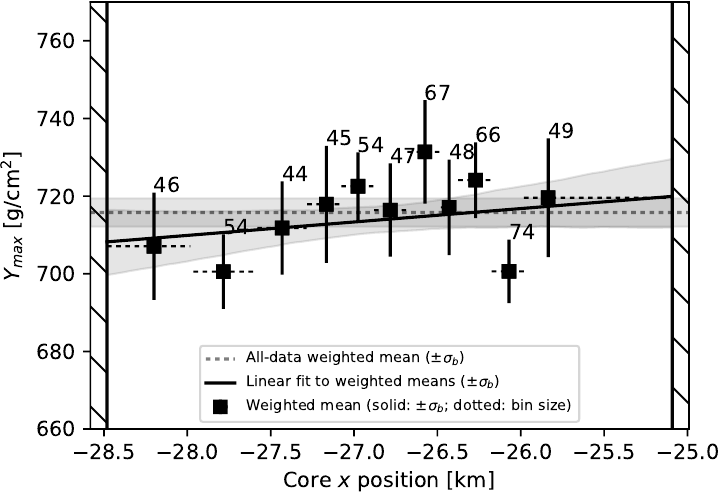}
 \hfill
    \includegraphics[width=0.48\textwidth]{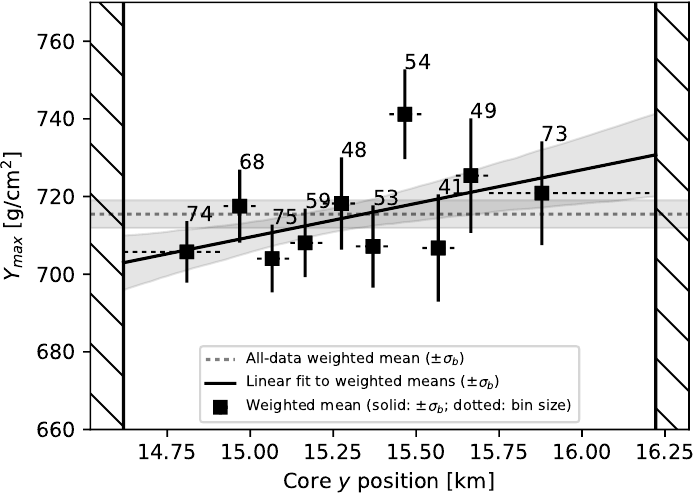}
    \caption{\label{fig:ResidualSyst} Relation between \Ymax [Eq.~(\ref{eq:ymax})] and the cosine of the zenith angle (top left), geomagnetic angle (top right), core position $x$ coordinate (bottom left), and core position $y$ coordinate (bottom right). Both core coordinates are relative to the center of the Auger array. The mean of $Y_\text{max}$ is shown in equally-spaced bins, or merged bins if containing less than $40$ showers (black squares). The number of events per bin is quoted next to each bin. The solid-line error bars show the uncertainties on the means, determined with bootstrap resampling. The dashed-line bars indicate the extent of each bin. Also shown are the mean of the entire data set (dashed line with $1\sigma$-confidence band) and a linear fit to the mean values (solid line with $1\sigma$-confidence band).}
  \end{figure*}

 We investigate the dependence of \Ymax as a function of shower zenith angle $\theta$, geomagnetic angle $\alpha$, and core position in $x$ and $y$ coordinates (relative to the center of the Auger array). The azimuth angle $\phi$ has been checked for a possible sinusoidal trend, but shows no variation that is not already explained by zenith or geomagnetic angle dependencies, hence is not shown. For each of these geometry parameters, event selection might be affected by the irregular AERA antenna spacing and various antenna hardware types used in AERA, despite the acceptance cuts that were implemented (see Sec.~\ref{sec:systematics}). We split up the \Ymax data set in regular bins for each of these variables and calculate the mean of \Ymax in each of these bins. The uncertainty on \Ymax combines the uncertainties of \Xmax and the energy dependence that was compensated for. The uncertainty on these mean values is determined by bootstrap resampling where we repeatedly sample $75$\% of the data and look at the variations in the mean. This procedure is done such that both the non-Gaussian distribution of $Y_\text{max}$, the uncertainties on $Y_\text{max}$, and statistical uncertainties from the limited number of showers can all be accounted for properly.

  Fig.~\ref{fig:ResidualSyst} shows the mean values as a function of the cosine of the zenith angle, geomagnetic angle, shower core $x$ coordinate, and shower core $y$ coordinate, respectively (black squares). The solid-line error bars show the uncertainties on the means from the bootstrap resampling. The dashed-line bars indicate the extent of each bin. Also shown are the mean of the entire data set (dashed line with $1\sigma$-confidence band) and a linear fit to the mean values (solid line with $1\sigma$-confidence band). The hatched regions indicate where no data is present, either from prior cuts or physical constraints.
  
  None of the trends show a significant deviation from the all-data mean values within their uncertainties and as such we can say that there are no significant residual biases that depend on these geometry parameters. In other words, any possible bias in \Ymax would need to be independent of geometry, which would be difficult to envision. 
  
  This is not to say that there is no residual bias; within the fit uncertainties there is the possibility for bias to exist. Hence, next we calculate the expected possible residual bias within the statistical uncertainties of our data set. We take as reference point the values of the geometry parameters $G \subset (\cos{\theta},\alpha,x,y)$ where the least bias is expected (i.e., where detector sensitivity is optimal). This will provide us with an estimate for possible bias in possibly less sensitive regimes. Using the linear fits $Y(G)$ we calculate the shift of each shower \Ymax value under the assumption that the \Ymax value at the expected $G$ value $G_\text{exp}$ is the true bias-free value. This expected value for zenith angle is $\theta_\text{exp}=55^\circ$, the highest zenith angle allowed in our data set and also where the footprints are largest and antenna sensitivity is excellent for the AERA antenna types. Hence, the acceptance would arguably be least affected there and any bias in less sensitive regimes we can determine w.r.t.\ this value. For the geomagnetic angle we choose $\alpha_\text{exp}=180^\circ$, where the geomagnetic radio emission is at a maximum. For the core position we pick $x_\text{exp}=-26.1$\,km and $y_\text{exp}=15.1$\,km, corresponding approximately to the center of the part of AERA with the densest antenna spacing (see Fig.~\ref{fig:AERAmap}).
  
  This shift in \Ymax is then given for each individual shower by:
  
    \begin{equation}
  \Delta Y_\text{max} = Y(G)-Y(G=G_\text{exp}).
  \end{equation}
  
  The bias for each geometry parameter $G$ is then given by the difference of the mean \Ymax of the showers in a certain energy bin, with and without the correction to the expected bias-free $G$:
  
  \begin{equation}\label{eq:residualBiasEq}
\text{Bias}_G          = \langle Y_\text{max} \rangle - \langle Y_\text{max}-\Delta Y_\text{max} \rangle .
  \end{equation}
  
  It should be noted that the biases from the geometry parameters are highly correlated quantities and as such cannot be summed in quadrature to get a total uncertainty. Hence, the extrema of the set of possible biases $\{\text{Bias}_G\}$ is used to determine an upper and lower limit on the possible residual bias:
      \begin{eqnarray}
      \text{Bias}^\text{up} &=& \text{Max}\left[ \{\text{Bias}_G\}  \right],\\
      \text{Bias}^\text{low} &=& \text{Min}\left[ \{\text{Bias}_G\}  \right]. \label{eq:resbiasuplow}
  \end{eqnarray}
  
  This then provides an estimation for the possible residual bias in \Ymax allowed within the statistical uncertainties of the shower data set, for each of the energy bins. From the definition of \Ymax [Eq.~(\ref{eq:ymax})] it follows that these $\langle Y_\textup{max} \rangle$ bias values apply also to $\langle X_\textup{max} \rangle$. The result of the $\text{Bias}^\text{up}(E)$ and $\text{Bias}^\text{low}(E)$ on $\langle X_\textup{max} \rangle$ are shown in Fig.~\ref{fig:TotalSyst}. The the possible bias does not seem to be dominated by any single effect and does not change significantly with energy. It indicates that within the statistical uncertainties with which we can constrain any bias we are potentially biasing our \Xmax values between $\pm7$\,g\,cm$^{-2}$. Table~\ref{tab:resbias} lists the contributions of the geometry parameters per energy bin. All values are well within the statistical uncertainty on $\langle X_\textup{max} \rangle$ itself and hence there is no hint of any significant residual bias in this analysis.
  
  \begin{table*}[!ht]
    \caption{\label{tab:resbias}%
    Table of possible residual biases [Eq.~(\ref{eq:residualBiasEq})] for geometry dependent parameters. Values are shown for each energy bin. The two right-most columns show the combined upper and lower limits by taking the extrema of the 4 parameters [Eq.~(\ref{eq:resbiasuplow})]. These values are also shown in Fig.~\ref{fig:TotalSyst}.
    }
    \begin{ruledtabular}

\begin{tabular}{l|c|c|c|c||c|c}
    lg($E\textup{[eV]}$) range & $\cos(\theta)$ [g\,cm$^{-2}$]& $\alpha$ [g\,cm$^{-2}$] & core $x$ [g\,cm$^{-2}$] & core $y$ [g\,cm$^{-2}$]& Lower limit [g\,cm$^{-2}$]& Upper limit [g\,cm$^{-2}$]\\ \hline
{[}$17.50$, $17.65$) &$4.4$ &$-3.3$ &$0.9$ &$-3.3$ &$-3.3$ &$4.4$\\[1mm]
{[}$17.65$, $17.80$) &$5.6$ &$-3.9$ &$1.2$ &$-3.0$ &$-3.9$ &$5.6$\\[1mm]
{[}$17.80$, $17.95$) &$4.9$ &$-4.3$ &$2.3$ &$-3.7$ &$-4.3$ &$4.9$\\[1mm]
{[}$17.95$, $18.10$) &$6.6$ &$-5.9$ &$2.8$ &$-4.0$ &$-5.9$ &$6.6$\\[1mm]
{[}$18.10$, $18.25$) &$4.7$ &$-6.7$ &$2.9$ &$-4.3$ &$-6.7$ &$4.7$\\[1mm]
{[}$18.25$, $\infty$) &$5.3$ &$-5.9$ &$3.8$ &$-1.0$ &$-5.9$ &$5.3$\\[1mm]
\end{tabular}
    \end{ruledtabular}
    \end{table*}
    
      \begin{table*}[!ht]
    \caption{\label{tab:xmaxmoments}%
    Table of the two moments of the \Xmax distribution for the six energy bins. Also listed are the ranges and mean energies for each energy bin and the number of showers in each bin. The two \Xmax moments are listed together with, in order, their $1\sigma$ statistical and systematic uncertainties as shown in Fig.~\ref{fig:elongation}.
    }
    \begin{ruledtabular}
    
    \begin{tabular}{l|c|r|lll|lll}
    lg($E\textup{[eV]}$) range&
    $\langle \textup{lg}(E\textup{[eV]}) \rangle$&
    $N$&
    \multicolumn{3}{l|}{$\langle X_\textup{max} \rangle$ $\pm$ stat. $\pm$ syst. [g\,cm$^{-2}$]}&
    \multicolumn{3}{l}{$\sigma(X_\textup{max})$ $\pm$ stat. $\pm$ syst. [g\,cm$^{-2}$]}\\
    \colrule%

    &&&&&&&&\\[-3mm]

{[}$17.50$, $17.65$) &$17.6$ &$167$ &$700.7$ &$\pm7.2$ &$_{-19.0}^{+11.7}$ &$67.1$ &$_{-12.8}^{+12.8}$ &$_{-0.4}^{+18.0}$\\[1mm]
{[}$17.65$, $17.80$) &$17.7$ &$150$ &$709.7$ &$\pm6.3$ &$_{-12.5}^{+11.4}$ &$55.2$ &$_{-8.9}^{+9.3}$ &$_{-0.3}^{+14.8}$\\[1mm]
{[}$17.80$, $17.95$) &$17.9$ &$127$ &$706.2$ &$\pm7.6$ &$_{-10.3}^{+9.9}$ &$64.9$ &$_{-10.1}^{+10.8}$ &$_{-0.7}^{+9.2}$\\[1mm]
{[}$17.95$, $18.10$) &$18.0$ &$74$ &$720.7$ &$\pm8.2$ &$_{-10.5}^{+10.2}$ &$52.7$ &$_{-8.6}^{+9.5}$ &$_{-0.6}^{+7.7}$\\[1mm]
{[}$18.10$, $18.25$) &$18.2$ &$43$ &$730.3$ &$\pm12.7$ &$_{-11.6}^{+8.3}$ &$68.3$ &$_{-11.6}^{+11.9}$ &$_{-0.3}^{+10.4}$\\[1mm]
{[}$18.25$, $\infty$) &$18.4$ &$33$ &$771.1$ &$\pm17.9$ &$_{-11.4}^{+10.1}$ &$78.3$ &$_{-16.0}^{+15.9}$ &$_{-0.3}^{+9.3}$\\[1mm]
    \end{tabular}
    \end{ruledtabular}
    \end{table*}
    
  Cross-checks have been done to test that additional artificially introduced biases (e.g., adding a linear \Xmax($\alpha$) dependence) can be recovered with this procedure to a degree that the $\text{Bias}^\text{up}(E)$ and $\text{Bias}^\text{low}(E)$ indeed account for the artificial bias. Additionally, also the effect on the median \Ymax versus the geometry parameters has been evaluated. Compared to using the mean, the median is less sensitive to the shape of the tail of a distribution (i.e., less sensitive to large outliers) and thus has an increased sensitivity to a more general shift of the distribution. Also here no significant trend was found and the allowed possible biases were similarly small.

\section{Calculation of the Second Moment of the $X_\text{max}$ Distribution}\label{app:sigmaxmax}
    The true distribution of $X_\textup{max}$ can be estimated from the width of the AERA \Xmax distribution by subtracting the effect of the method resolution. Since the uncertainty of the method is not a perfect Gaussian distribution one can't simply calculate this with 
    \begin{equation}\label{eq:NaiveSecondMoment}
    \sigma_\textup{true}=\sqrt{\sigma_\textup{measured}^2-\sigma_\textup{method}^2} ,
    \end{equation}
    because there is no simple single value for $\sigma_\textup{method}$ for a distribution of arbitrary shape. To account for this, a bootstrapping procedure is applied, where the uncertainty of the method is repeatedly randomly sampled and subtracted in quadrature from the total measured width of the AERA \Xmax distribution. Assuming that the spread in the AERA \Xmax distribution due to the method resolution and the true spread are uncorrelated, the distribution of the average true spread (i.e., the intrinsic spread in \Xmax due to shower-to-shower fluctuations) is estimated by
  \begin{equation}
  \hspace*{-0.2cm}
  \sigma(X_\textup{max}) \equiv B_N\left( \sqrt{\textup{Var}\left(  \varphi_{75} \left(X_\textup{max}\right)\right) - \varphi_{75|1} \left( \left(\delta_{X_\textup{max}}\right)^2\right)} \right ). \label{eq:sigmaXmaxPhysics} \\
  \end{equation}
  
  Note that this equation still closely mirrors Eq.~\ref{eq:NaiveSecondMoment}, but with some extra steps. $B_N(x)$ we define as the distribution given by performing $N$ bootstrapping iterations on the argument $x$, $\varphi_{75}(y)$ is defined as the function that samples $75$\% of a data series $y$ at random for the bootstrapping of $B_N$, and $\varphi_{75|1}(y)$ the function that selects one value at random from $\varphi_{75}(y)$ and returns $\delta_y$, the uncertainty on $y$. The first term in the square root then represents the width of the AERA $X_\textup{max}$ distribution of the showers and the second term is the $X_\textup{max}$ uncertainty estimation of the method. The number of iterations $N$ for the bootstrapping is set at $10000$ to sample the whole distribution sufficiently. 

  The mean and width of the $\sigma(X_\textup{max})$ distribution $B_N$ can now be calculated, but also this distribution is not necessarily a Gaussian distribution. We take the mean and the quantile region equivalent to the probability contained in a $1\sigma$ standard deviation of a Gaussian (i.e., the region between the $15.87$\% and $84.13$\% quantiles) will be quoted as the mean and (asymmetric) uncertainty on $\sigma(X_\textup{max})$, respectively.

\section{Tabulated $X_\text{max}$ Moments}\label{app:tabulated}
  
 Table~\ref{tab:xmaxmoments} lists the values of the two central moments of the \Xmax distribution and their uncertainties for six energy bins.



\end{document}